\documentclass[onecolumn,showpacs,showkeys,preprintnumbers,amsmath,amssymb]{revtex4}
\newcommand{\old}[1]{
} 

\usepackage[final]{graphicx}
\usepackage{dcolumn}
\usepackage{bm}
\usepackage{slashed}
\usepackage{multirow}
\begin{document}

\title{Universal correlations in pion-less EFT with the Resonating Group Method:\\ three
  and four nucleons}
\author{Johannes Kirscher}\email[Corresponding author; electronic
address: ]{kirscher@gwu.edu} \author{Harald W.~Grie\ss hammer}
\affiliation{Center for Nuclear Studies, Department of Physics,
The George Washington University, Washington, DC 20052, USA}
\author{Deepshikha Shukla}\affiliation{Department of Physics and Astronomy,
University of North Carolina, Chapel Hill, NC 27599-3255, USA}
\author{Hartmut M.~Hofmann} \affiliation{Institut
  f\"ur Theoretische Physik III, Department f\"ur Physik, Universit\"at
  Erlangen-N\"urnberg, D-91058 Erlangen, Germany}

\date{\today}


\begin{abstract}
  The Effective Field Theory ``without pions'' at next-to-leading order is
  used to analyze universal bound state and scattering properties of the 3-
  and 4-nucleon system.  Results of a variety of phase shift equivalent
  nuclear potentials are presented for bound state properties of $^3$H
  and $^4$He, and for the singlet S-wave $^3$He-neutron scattering length
  $a_0(^3\textrm{{\small He-n}})$. The calculations are performed with the
  Refined Resonating Group Method and include a full treatment of the Coulomb
  interaction and the leading-order 3-nucleon interaction. The results compare favorably with
  data and values from AV18(+UIX) model calculations.  A new correlation between
  $a_0(^3\textrm{{\small He-n}})$ and the $^3$H binding energy is found.
  Furthermore, we confirm at next-to-leading order the correlations,
  already found at leading-order, between the $^3$H
  binding energy and the $^3$H charge radius, and the Tjon line. With the $^3$H binding energy as input, we get
  predictions of the Effective Field Theory ``without pions'' at next-to-leading order
  for the root mean square charge radius of $^3$H of $(1.6\pm 0.2)~\textrm{fm}$, for
  the $^4$He binding energy of $(28\pm 2.5)~\textrm{MeV}$, and for
  $\textrm{Re}\lbrace a_0(^3\textrm{{\small He-n}})\rbrace$ of $\left(7.5\pm 0.6\right)~$fm.
  Including the Coulomb interaction, the splitting in binding energy between $^3$H and
  $^3$He is found to be $\left( 0.66\pm 0.03\right)~\textrm{MeV}$. The discrepancy to data
  of $\left( 0.10\mp 0.03\right)~\textrm{MeV}$ is model independently attributed to higher order
  charge independence breaking interactions. We also demonstrate that
  different results for the same observable stem from higher order effects, and carefully assess that
  numerical uncertainties are negligible.
  Our results demonstrate the convergence and usefulness of the pion-less theory at
  next-to-leading order in the $^4$He channel. We conclude
  that no 4-nucleon interaction is needed to renormalize the theory at
  next-to-leading order in the 4-nucleon sector.
\end{abstract}

\maketitle
\section{Introduction}\label{ch.intro}

Nuclear processes, both with and without external probes, at energies well
below the pion production threshold are naturally described by the Effective
Field Theory ``without pions'' EFT$_\slashed{\pi}$ (see \textit{e.g.}
\cite{eft-rev-vK,eft-rev-bb,eft-rev,eft-rev-platter,brat-hammer} for reviews). It shares with QCD the symmetries
and effective low-energy degrees of freedom and provides a simple, systematic,
and model independent approach appropriate for systems at very low energies.
Like all Effective Field Theories, it allows for a systematic expansion of the
scattering amplitude in a small, dimensionless parameter to attain the desired
level of accuracy in observables and therefore provides reliable \emph{a priori}
error estimates. Its simple structure makes it also easier to handle
than pionful EFT~\cite{mod-rev-nucl-epel,xpt-epelbaum,epel-chipt-rev}.

EFT$_\slashed{\pi}$ has successfully been applied to many reactions of two and
three nucleons with and without electro-weak currents. While calculations at
fourth order have been employed to yield very accurate results in those
systems (see \textit{e.g.}~\cite{rupak-npdgamma}), higher order calculations involving more than
three nucleons are still in their infancy.  This work supplements the few,
already existing leading-order (LO) calculations in the bound four-nucleon
system~\cite{platter-tjon,vkol-ncsm-li} by a next-to-leading order (NLO)
analysis including the Coulomb interaction and the leading-order 3-nucleon interaction.
As first observable in the four-nucleon scattering system, it adds the
singlet S-wave \mbox{3-Helium-neutron} scattering length $a_0(^3\textrm{{\small He-n}})$.
By that, it addresses the question whether EFT$_\slashed{\pi}$ still converges
for the $\alpha$-particle.

Our investigation has three central goals: (i) to support evidence that the four-nucleon
bound- and scattering system can be described by EFT$_\slashed{\pi}$; (ii) to provide
strong evidence that no four-nucleon-interaction is necessary to renormalize EFT$_\slashed{\pi}$ at NLO; and most importantly
(iii) to clearly demonstrate the feasibility of the variational approach of the "RRGM" (RRGM) for
 EFT$_\slashed{\pi}$ in addressing these goals also for future studies involving heavier systems.

EFT$_\slashed{\pi}$ systematically expands observables in powers of the
typical low-momentum scale $p_\textrm{typ}$ of the process
considered, measured in units of the scale
$\Lambda_{\textrm{EFT}_\slashed{\pi}}$ at which the pion can be resolved as a
dynamical degree of freedom, set by the pion mass $m_\pi$. Equivalently, the
expansion parameter is given by the ratio of the range $r_\pi\approx 1.4$~fm
of the one-pion-exchange over typical resolutions or sizes of the process under
consideration. The \mbox{4-Helium} binding energy is $28.5\,$MeV and its typical
size of about $1.6\,$fm is comparable with the approximate range of the
one-pion-exchange. A systematic expansion in the four-nucleon system could
therefore converge very slowly at best. However, Platter
\textit{et al.}~\cite{platter-tjon} showed that EFT$_\slashed{\pi}$ at LO can explain
with good accuracy the correlation between the \mbox{4-Helium} and triton binding
energies, the famous Tjon line~\cite{tjon}. One focus of this article
is to test whether or not EFT$_\slashed{\pi}$ at NLO shows a convergence
order-by-order, and converges to experimental values.

For this purpose, we investigate correlations between three- and four-nucleon
observables. They correspond to universality classes of nuclear interaction models,
all resulting in on-shell (phase shift) equivalent 2-nucleon potentials
which share the same two-body scattering lengths $a_{s,t}$ in the singlet
and triplet $S$-wave. The elements of a class, parameterized by $a_{s,t}$,
therefore fully describe the nucleon-nucleon (NN)
system in the zero energy limit. In contradistinction, nonuniversal parameters
like the effective range only add perturbative corrections and correspond to
higher order interactions in the EFT power-counting. Their contribution to an
observable vanishes in the zero energy limit.  The empirically found
Phillips~\cite{phillips} line, the Efimov spectrum of three-body bound
states and the Tjon line~\cite{tjon} are
examples for universal properties of NN potentials of the same universality
class defined by $a_{s,t}$ (for reviews see \textit{e.g.}~\cite{eft-rev-platter,efi-plat-hamm,brat-hammer}).
The Phillips line is a correlation between the triton binding
energy $B(t)$ and the neutron-deuteron doublet scattering length and can be explained as a
result of the fact that on-shell equivalent 2-nucleon potentials have in
general different off-shell behavior. Such a dependence of low-energy
observables on details of short-distance 2-nucleon physics is however
unphysical. This renormalization defect can be corrected by including
a three-body interaction which eliminates the dependence on the NN
potential. The datum on the Phillips line is then reproduced at LO by
fitting the strength of the three-body interaction~\cite{efimov-spect,bedaque-tni}. In the
zero energy limit, the 3-nucleon system is hence described by three
universal parameters: $a_{s,t}$ and one three-body low-energy datum.
Different values for this low-energy datum result in a different strength
of the 3-nucleon interaction. As the latter is however also dependent on
the NN potential used, one can use an alternative approach to explore the
dependence of few-nucleon observables on the additional three-body input:
Set the three-body interaction to zero and map out the dependence of
observables on the regularization procedure used for the 2-nucleon
potential. The results are correlations between few-body observables.
In this work, we employ all
three procedures to map out correlations: different values of the regulating cutoff, a
variation of unresolved short-distance observables by allowing for
non-zero, higher order P-wave and SD interactions, and a variation of the
coupling strength of the three-body counterterm with fixed 2-nucleon
interaction.

One might be tempted to choose form and value of the cutoff for the
2-nucleon potential such that the triton binding energy is reproduced
``exactly'', and then perform all calculations with such a highly fine-tuned
potential, see e.g.~\cite{kharchenko}. However, such an approach would deprive
one of a simple procedure to find an estimate of the theoretical uncertainty
of the calculation.  Including electromagnetic interactions will also be more
cumbersome, as these form-factors have to be gauged and additional, transverse
currents need to be constructed. On top of that, it is not clear that one can
find such a 2-nucleon potential to eliminate any 3-nucleon interaction, or
that its choice would be unique~\footnote{We thank W.~Gl\"ockle and
D.~R.~Phillips for remarks on this point, which helped to sharpen our view
on this issue.}. The numerical advantage of not having to deal with
3-nucleon interactions in some systems can in general be balanced by the
problem to find the ``right'' 2-nucleon potential. In any case, such
additional effort does not increase the reliability of the calculation. We
therefore keep the cutoff arbitrary to map out correlations and
error estimates.

Adding one more nucleon, the Tjon line of a one-parameter correlation
between the binding energies of the triton and \mbox{4-Helium} supports the
assertion that no additional universal parameter is necessary to classify the
four-nucleon system. Its emergence in EFT$_\slashed{\pi}$ at LO was
demonstrated by Platter \textit{et al.}~\cite{platter-tjon}.  In order to extend this
finding to NLO, this work investigates not only the dependence of the triton
charge radius, the splitting in the trinucleon binding energies,
and the \mbox{4-Helium} binding energy on the triton binding energy,
but also finds a similar connection between the real part of $a_0(^3\textrm{{\small He-n}})$
and the triton binding energy.  Therefore, it supports the expectation, based
on na\"ive dimensional analysis, that no four-body interaction is necessary
to renormalize the theory at NLO in the four-body sector. Up to this level of accuracy, observables in the
four-nucleon system are universal consequences of the two- and three-body
dynamics of its constituents. For all observables we consider, namely the charge radius
of the triton, the trinucleon binding energy splitting, the 4-Helium binding energy $B(\alpha)$ and
the scattering length $a_0(^3\textrm{{\small He-n}})$,
we find predictions consistent with experiment within error margins when the 3-nucleon interaction
strength is fitted to $B(t)$.

For this work, the Refined Resonating Group Method~\cite{hmh-rrgm} is
employed. This variational method numerically solves the Schr\"odinger
equation in coordinate representation. Within a model space spanned by
Gaussian type wave functions, minimization of a respective functional yields
scattering and bound state observables for a potential parameterized in terms
of Gaussians.
With the RRGM, speedy scattering calculations are feasible in the $A=4$
system and even beyond, because of the relatively simple structure of the NLO
EFT$_\slashed{\pi}$ potentials. To study the feasibility of using
EFT$_\slashed{\pi}$ in the RRGM is the other focus of this article.
In the course of this investigation, it is also imperative to carefully assess that the spread of
results does not come from numerical inaccuracies.

The article is organized as follows: First, a brief overview of
EFT$_\slashed{\pi}$ is given as a low-energy theory which allows for a
systematic improvement of the interaction amongst nucleons up to the desired
accuracy.  An introduction to the calculational tool of the RRGM in
sect.~\ref{ch.rrgm} is followed by the derivation, structure, and fit of the NN
potential in sect.~\ref{ch.potfit}. The results are presented in
sect.~\ref{ch.res}, followed by a concluding section.
An appendix addresses the numerical accuracy and computational costs
of the calculations.
\section{Pion-less theory}\label{ch.piless}

In this section, the theoretical framework of an Effective Field Theory for nucleons
without pions is recapitulated. For details, the reader is referred to the
exemplary reviews~\cite{eft-rev,eft-rev-phill}.

The nuclear potential is derived from a Lagrangean (see
\textit{e.g.}~\cite{eft-ord-ray-vkol}) with contact interactions which are
momentum independent at LO and momentum dependent at NLO:
\begin{small}
  \begin{eqnarray}\label{eq.nlo-lagrangean}
    \mathcal{L}_{\slashed{\pi},NN}&=&N^\dagger\left(i\partial_0+\frac{\mathbf{\nabla}^2}{2M}\right)N+C^{\textrm{LO}}_1(N^\dagger N)(N^\dagger N)+
    C^{\textrm{LO}}_2(N^\dagger\sigma_i N)(N^\dagger\sigma_i N)\nonumber\\
    &&- \frac{1}{2} C^{\textrm{NLO}}_1  \left[  ( N^\dagger  \partial_i N )^2  +
      \left(( \partial_i N^\dagger ) N \right)^2  \right]\nonumber\\
    &&   -  \left( C^{\textrm{NLO}}_1 - \frac{1}{4} C^{\textrm{NLO}}_2 \right) ( N^\dagger \partial_i N ) \;\Big[( \partial_i N^\dagger ) N \Big]+
    \frac{1}{8} C^{\textrm{NLO}}_2  ( N^\dagger N ) \left[ N^\dagger\partial_i^2 N + \partial_i^2 N^\dagger N \right]\nonumber\\
    &&  - \frac{i}{8} C^{\textrm{NLO}}_5  \epsilon_{ijk} \bigg\{\left[ ( N^\dagger \partial_i N ) \big[( \partial_j N^\dagger ) \sigma_k N \big] +
      \big[ ( \partial_i N^\dagger ) N \big]  ( N^\dagger\sigma_j\partial_k N ) \right]\nonumber\\
    &&-( N^\dagger N )\Big[ ( \partial_i N^\dagger ) \sigma_j\partial_k N \Big]  +   ( N^\dagger \sigma_i N )\Big[ ( \partial_j N^\dagger )\partial_k N \Big] \bigg\}+ \frac{1}{4} \Big[ \left( C^{\textrm{NLO}}_6 + \frac{1}{4} C^{\textrm{NLO}}_7  \right)    \left( \delta_{i k}\delta_{j l}+ \delta_{i l} \delta_{k j} \right)\nonumber\\
    && +\left( 2 C^{\textrm{NLO}}_3 + \frac{1}{2} C^{\textrm{NLO}}_4 \right) \delta_{i j} \delta_{k l} \Big]\,\Big[  \big[ (\partial_i \partial_j N^\dagger)\sigma_k  N \big]+
    ( N^\dagger \sigma_k \partial_i \partial_j N ) \Big] ( N^\dagger\sigma_l N )\nonumber\\
    &&  -  \frac{1}{2} \Big[ C^{\textrm{NLO}}_{6} \left( \delta_{i k} \delta_{jl}+ \delta_{i l} \delta_{k j} \right) +
    C^{\textrm{NLO}}_4 \delta_{i j} \delta_{k l} \Big]( N^\dagger \sigma_k \partial_i N ) \big[( \partial_j N^\dagger ) \sigma_l N \big]\nonumber\\
    &&- \frac{1}{8} \left( \frac{1}{2}  C^{\textrm{NLO}}_7\left( \delta_{i k} \delta_{j l}+ \delta_{i l} \delta_{k j}\right) -
      ( 4 C^{\textrm{NLO}}_3 - 3 C^{\textrm{NLO}}_4 ) \delta_{i j} \delta_{k l} \right)\nonumber\\
    &&\left[  ( \partial_i N^\dagger \sigma_k\partial_j N ) +
      ( \partial_j N^\dagger\sigma_k \partial_i N ) \right] ( N^\dagger \sigma_l N )\;\;\;.
  \end{eqnarray}
\end{small}
The only effective low-energy degree of freedom is the nonrelativistic fermion
iso-doublet $N=\left(\begin{array}{c}p\\n\end{array}\right)$ of Weyl spinors
$p$ and $n$ for the proton and neutron, respectively. In eqs.~(\ref{eq.nlo-lagrangean},\ref{eq.nlo-lagrangean-3nf}),
Einstein's summation convention is understood, \textit{i.e.}, a sum from 1 to 3 over repeated Arabic indices. The
Lagrangean is an isoscalar at this order, with neutrons and protons having the same mass
$M$.  The coupling constants $C_i^{\textrm{(N)LO}}$ are referred to as low-energy
constants (LECs) or counterterms, and $\sigma_i$ are the Pauli spin matrices.
The power-counting in EFT$_\slashed{\pi}$ at NLO (see \textit{e.g.}~\cite{eft-rev-bb}) results in four independent LECs,
with zero P-wave and SD-transition amplitudes
providing the five constraining equations (see the discussion of eq.~(\ref{eq.p-wave-hard}) for
how this constraint is implemented).

Within the EFT framework, it was found~\cite{bedaque-tni-boson,bedaque-tni,bedaque-hgrie-tni,ha-meh-nd-scatt,platter-NNLO,platt-phill-NNLO}
that exactly one 3-nucleon contact interaction (3NI) is necessary to renormalize
the $A=3$ system in the doublet-S channel at LO and NLO.
In our NLO calculation, this three-body counterterm has the form
\begin{equation}\label{eq.nlo-lagrangean-3nf}
 \mathcal{L}_{\slashed{\pi},3N}=C^{\textrm{LO}}_{3\textrm{NI}}(N^\dagger N)(N^\dagger\tau_i N)(N^\dagger\tau_i N)\;\;\;.
\end{equation}
Other forms are identical after Fiertz-transformations~\cite{bedaque-hgrie-tni}.
Once the 2-nucleon parameters in eq.~(\ref{eq.nlo-lagrangean}) are fixed, the
universal correlation lines between three- and four-nucleon observables, discussed in sect.~\ref{ch.res},
are parameterized by the 3-nucleon coupling $C^{\textrm{LO}}_{3\textrm{NI}}$.

The amplitudes derived from the Lagrangean in eq.~(\ref{eq.nlo-lagrangean}) are given
as an expansion in the dimensionless parameter $Q\sim p_\text{typ}/{\Lambda_b}$, where $p_\text{typ}$
is a typical low-momentum scale of the system, and a rough estimate for the breakdown scale
$\Lambda_b$ of the theory is the pion mass.
In the neutron-proton system at center of mass energies of less than the deuteron binding energy $B(d)$,
$p_\text{typ}$ is set by the binding momentum, $\sqrt{MB(d)}$, which leads to an expansion parameter of $Q\lesssim \frac{1}{3}$.
A calculation at N$^n$LO is then expected to be accurate up to perturbative corrections of order $Q^{n+1}$. For energies below $B(d)$, the
parameter $Q$ is found to be approximately constant (\textit{e.g.}~\cite{lepage}) but increases for higher energies. This increase leads eventually
to a breakdown of the perturbative expansion.
In practice, a cornucopia of EFT$_\slashed{\pi}$ calculations~\cite{eft-rev,rupak-npdgamma,bedaque-hgrie-tni,christl-ddis,hgrie-zpara}
to higher orders has shown that the expansion converges somewhat
beyond the pion mass and that $\frac{1}{5}\lesssim Q\lesssim\frac{1}{3}$.

The NN interaction is understood as an effective potential following
Weinberg's original definition~\cite{wein-chi-nucl}. In the pion-less theory,
the potential reduces to tree diagrams of two in- or out-going nucleons, with
vertices from eq.~(\ref{eq.nlo-lagrangean}). While not necessary from the EFT
standpoint, it is convenient for this work to insert the full NLO potential in
this form into the Schr\"odinger equation instead of treating the non-leading
terms in perturbation theory. This course of action has been pursued regularly
and includes some contributions which are formally of higher order in the
EFT$_\slashed{\pi}$ power-counting but does not increase the accuracy of the
result. The re-summation does however lead to spurious bound states in the
two-particle sector which can impact three-particle observables, see
\textit{e.g.}~\cite{hgrie-zpara}.

This method also allows us to briefly comment on how our results would change
in an alternative power-counting in EFT$_\slashed{\pi}$ proposed by Beane and
Savage~\cite{beane-savage-rearr}, where both scattering lengths and
effective ranges count as $1/p_{\text{typ}}\sim Q^{-1}$.
In that case, effective-range corrections must be re-summed, and some
combinations of the NN interactions $C_{1-7}^\text{NLO}$ in
eq.~(\ref{eq.nlo-lagrangean}) are promoted to LO. One would now be compelled
to iterate the potential in the Schr\"odinger equation. As one performs
technically the same steps as above, one arrives at the same amplitude. Our
results can thus also be interpreted as LO calculations in the alternative
formulation. The difference between this alternative and the approach taken
here is therefore only in the question whether re-summing effective range
contributions is \emph{optional} or \emph{mandatory}. If the effective ranges
can but do not have to be included as NLO corrections, one expects that
observables change only by parametrically small amounts when one calculates
first with zero effective ranges (i.e.~keeping $C_{1,2}^\text{LO}$ only), and
then adds the supposed NLO terms $C_{1-7}^\text{NLO}$. Our results for the
triton charge radius in sect.~\ref{ch.res.friar-line} and for the Tjon line in
sect.~\ref{ch.res.tjon} confirm this assumption in the three- and four-nucleon
system. From that perspective, we therefore see no reason to make the
effective-range resummation mandatory.

As pointed out in~\cite{wign-bound-phil}, for a short ranged potential like this,
the Wigner bound can potentially constrain the value of the effective ranges: When the scattering length is positive,
an energy-independent potential exists only below a cutoff-dependent upper
bound for the effective range. For high enough cutoffs, the physically
observed effective range in the $^3S_1$ channel will exceed this Wigner
bound. However, we will demonstrate in sect.~\ref{ch.potfit} that for our choice of cutoffs,
the problem does not arise.

The derivation of the explicit form of the potential in coordinate space from
eq.~(\ref{eq.nlo-lagrangean}) is postponed to sect.~\ref{ch.potfit} in favor of
first introducing the numerical method used, as it motivates also the choice
of operator structure for the potential.

\section{Resonating Group Method}\label{ch.rrgm}

We employ the variational method of the Refined Resonating Group Method (RRGM)
\cite{hmh-rrgm} to solve the Schr\"odinger equation in coordinate space.
In this section, the method is introduced. Factors determining the numerical
stability of EFT$_\slashed{\pi}$ calculations with the RRGM are addressed in app.~\ref{ch.rrgm.numstab}.

The RRGM uses a Gaussian basis to span the variational space and a Gaussian expansion of the
radial dependences of the nuclear potential which allow for an analytic
calculation of the Hamilton matrix. As in every variational approach, the
basis is incomplete. Care has to be taken to avoid linearly dependent basis
vectors. The computer time for the
calculation depends on the dimension of the variational space, the number
of Gaussians needed for an accurate fit of the radial functions of the
potential, and its operator structure. A Gaussian regulator for the contact interactions leads directly
to a Gaussian radial dependence of the potential and makes the RRGM an
efficient tool to analyze the few-nucleon sector. This section describes those
aspects of the method which are necessary to understand the choices made for
the form of the regulator function, for the range of its cutoff values, and
for the set of operators which constitute the potential.

To determine the bound state wave function, the Ritz functional is minimized
in a model space spanned by vectors of the form
  \begin{equation}\label{eq.bsbv}
    \psi_{\textrm{BS}}^{J^\pi}\left(\vec{\rho}_m,\vec{s}_m\right)=
\mathcal{A}\left\lbrace\sum_{d,i,j}c_{dij}\Bigg[\Big[\prod_{k=1}^{N-1}e^{-\gamma_{dk}\vec{\rho}_k^2}
\mathcal{Y}_{l_{ki}}\left(\vec{\rho}_k\right)\Big]^{L_i}\otimes\Xi^{S_j}\Bigg]^J\cdot\Upsilon\right\rbrace\;\;\;\;.
  \end{equation}
The system consists of $N$ particles.
To each of the $N-1$ Jacobi coordinates $\vec{\rho}_k$, one assigns a set of different width parameters
$\gamma_{dk}$ and of different angular momenta $l_{ki}$
represented by solid spherical harmonics $\mathcal{Y}_{lm}$~\cite{edmonds}. The antisymmetrizer is denoted by
$\mathcal{A}$, and the square brackets indicate angular momentum couplings,
with the orbital part $L_i$ being combined with the spin $S_j$ to the total
angular momentum $J$. The label $d$ distinguishes different sets of width
parameters $\gamma$, while $i$ and $j$ label sets of orbital- and spin angular momentum coupling schemes.
The spin function $\Xi$ is constructed as a sum of products of the single particle
spin functions $\vec{s}$.
The isospin function $\Upsilon$ is built
analogously from single particle isospin functions and distinguishes between
neutrons and protons.  The superposition coefficients $c_{dij}$ are
determined by minimizing the Ritz functional. The magnetic quantum numbers of
the spherical tensors are not explicit in the above equation where they are not needed to label a
specific basis vector. Since one deals only with reduced matrix elements,
even the magnetic quantum number corresponding to the total angular momentum
$J$ is of no significance.

For the scattering state, the Kohn-Hulth\'en variational principle is used with
the following ansatz for the wave function:
  \begin{eqnarray}\label{eq.ssbv}
    \psi_{\textrm{SS},\lambda}^{J^\pi}&=&
\mathcal{A}\Bigg\lbrace\sum_j^{n_k}\Bigg[\frac{1}{R_j}Y_{L_j}(\hat{\vec{R}}_j)\otimes\Big[\psi_j^{J_1^{\pi_1}}\otimes \psi_j^{J_2^{\pi_2}}\Big]^{S_{c_j}}\Bigg]^J\cdot\nonumber\\
&&\left(\delta_{\lambda j}F_{L_j}(R_j)+a_{\lambda j}\tilde{G}_{L_j}(R_j)+\sum_mb_{\lambda jm}R_j^{L_j+1}e^{-\omega_{jm}\vec{R}_j^2}\right)\Bigg\rbrace\;\;\;\;.
  \end{eqnarray}
The two fragments are represented by bound state wave functions
$\psi_j^{J_{1,2}^{\pi_{1,2}}}$, determined by the aforementioned Ritz
minimization and built from vectors as in eq.~(\ref{eq.bsbv}). The scattered
fragments are separated by $\vec{R}$, and the orbital angular
momentum between the two fragments is carried by a spherical harmonic $Y_L$.
$\lambda$ specifies the boundary condition which allows regular Coulomb waves
$F_L$~\cite{newton} only for the channel $j=\lambda$. The reactance
coefficients $a_{\lambda j}$ and the
$b_{\lambda jm}$ are determined by minimizing the Kohn-Hulth\'en functional,
taking into account $n_k$ channels.  The additional set of variational
parameters $b_{\lambda jm}$ is necessary to approximate the wave function in
the interaction region. The irregular Coulomb functions are regularized with a
polynomial weighted by an exponential and renamed $\tilde{G}_L$.  To become more familiar with the
terminology and as a precursor to the calculation presented in
sect.~\ref{ch.res}, consider a neutron scattered off a \mbox{3-Helium} nucleus.  In
that case, $n_k=654$ channels were included, the fragment wave function of the
neutron is $\psi_{(n)}^{\frac{1}{2}^+}=1$, and that of the \mbox{3-Helium}
$\psi_{(^3\textrm{He})}^{\frac{1}{2}^+}$. The latter consisted of 224 terms, namely
78 $(L_1=0,L_2=0)$ components, called $SS-$configuration, 82 $SD-$, 45 $DD-$, and 21 $PP-$configurations.
An example of an SD-configuration might be given by the following parameters: The angular momenta $l_{1}=0,l_{2}=2$
on the two Jacobi coordinates couple to a total orbital angular momentum $L=2$.
This mandates the individual spins of the
nucleons to be aligned, \textit{i.e.}, total spin $S=\frac{3}{2}$, so that
finally a part of the triton system with $J_2^\pi=\frac{1}{2}^+$ is formed.  The
two total angular momenta of the fragments,
$J_{1,2}^\pi=\frac{1}{2}^+$, can be coupled to channel spins, $S_c\in\lbrace 0,1\rbrace$.
However, only $S_c=0$ has to be included for the
$J^\pi=0^+$-channel, which in turn dictates that the orbital angular momentum
between the fragments is zero, $L_j=0$.  The width parameter $\omega_{jm}$ for
the inter-fragment wavefunction plus two widths $\gamma_{dk}$ for the \mbox{3-Helium}
bound state component complete the specification of the basis vector.  In the
following, the model space is defined to be a vector space spanned by basis
vectors $\psi_{\textrm{BS}/\textrm{SS}}$ as given in eq.~(\ref{eq.bsbv}) or eq.~(\ref{eq.ssbv}).

Once the potential is written in terms of spherical tensor operators and its
radial dependences are expressed in Gaussian functions, all the coordinate
space matrix elements which are needed to minimize the respective functional
can be cast into the form
\begin{equation}\label{eq.come}
  I=\int d^3\rho_1\ldots d^3\rho_{n_k-1}
  \exp\left[-\sum_{\nu\nu'}^{n_k-1}C^{\nu\nu'}\vec{\rho}_\nu\cdot \vec{\rho}_{\nu'}+
    \sum_{\nu}^{n_l}\vec{\xi}^\nu\cdot\vec{\rho}_\nu\right]\;\;.
\end{equation}
The matrix $C^{\nu\nu'}$ transforms
the Jacobi coordinates in the in-coming channel and the relative
coordinates of the radial dependences
in the potential operator to the coordinates in the out-going channel.
Furthermore, it takes into account that the antisymmetrizer permutes single
particle coordinates. The vector $\vec{\xi}^\nu$ is related to the
generating function of spherical harmonics. Its dimension is $n_l$, namely the
number of spherical harmonics in the matrix element $I$.  The integrals $I$
can hence be evaluated analytically for all operators of the EFT$_\slashed{\pi}$
potentials and the Coulomb interaction.

For an accurate description of the scattering state, so-called
distortion channels have to be added to the physical channels in
eq.~(\ref{eq.ssbv}). They increase the variational space to
allow for a more accurate description of the wave function in the interaction
region. There, a separation into two bound fragments as mimicked by the physical
channels does not resemble, \textit{e.g.} three- or four-body breakup states.
Those channels do not have an asymptotic tail,
\textit{i.e.}, only the square integrable terms with coefficients $b_{\lambda jm}$
in eq.~(\ref{eq.ssbv}) constitute the relative part of their wave
function.  Furthermore, their fragment functions are not restricted to
describe bound states, but only to have the correct quantum numbers.  That
means it suffices to specify a single set $\lbrace l_{ki},\gamma_{dk},S_j\rbrace$,
instead of superimposing multiple sets to a
bound state.  In the calculations, all but one of the components of each of
the physical channels were used for this purpose, \textit{e.g.}, a physical
\mbox{3-Helium-neutron} channel, with 224 basis vectors for the \mbox{3-Helium} fragment,
yields 223 configurations.  The number of distortion channels, and by that the
dimension of the model space, is ultimately determined by the number of
different $\omega_{jm}$ used in eq.~(\ref{eq.ssbv}).  With the number of
physical channels fixed, more distortion channels were added to obtain a
converged result for the observables.

In view of the error analyses performed in sect.~\ref{ch.res}, it is imperative to demonstrate
that purely numerical inaccuracies of the variational method are no significant source of
error.
In app.~\ref{ch.rrgm.numstab}, we demonstrate that differences in calculations of
the same observable do indeed not originate from numerical inaccuracies.

\section{Two-nucleon potential and parameter determination}\label{ch.potfit}

In this section, the potential in coordinate representation is derived from
the Lagrangean in eq.~(\ref{eq.nlo-lagrangean}), following the lines of
\cite{eft-ord-ray-vkol} but choosing a different operator structure which
simplifies the implementation into the RRGM. In the second part, the fitting
procedure of the low-energy constants (LEC) and the experimental input is
explained.

In momentum representation, the two-body potential for an $A$-nucleon system
following from the nine four-nucleon contact interactions of eq.~(\ref{eq.nlo-lagrangean}) is
  \begin{eqnarray}\label{eq.pot_mom}
    V^{(NLO)}_{\textrm{EFT}_\slashed{\pi},NN}=&\sum\limits_{i<j}^A\Big(C^{\textrm{LO}}_{1}+C^{\textrm{LO}}_{2}\,\vec{\sigma}_i\cdot\vec{\sigma}_j+C^{\textrm{NLO}}_1\vec{q}^2+
   C^{\textrm{NLO}}_2\vec{k}^2+\vec{\sigma}_i\cdot\vec{\sigma}_j\left(C^{\textrm{NLO}}_3\vec{q}^2+C^{\textrm{NLO}}_4\vec{k}^2\right)\nonumber\\
    &+iC^{\textrm{NLO}}_5\frac{\left(\vec{\sigma}_i+\vec{\sigma}_j\right)}{2}\cdot\vec{q}\times\vec{k}+C^{\textrm{NLO}}_6\vec{q}\cdot\vec{\sigma}_i\vec{q}\cdot\vec{\sigma}_j+
   C^{\textrm{NLO}}_7\vec{k}\cdot\vec{\sigma}_i\vec{k}\cdot\vec{\sigma}_j\Big)\;\;\;,
  \end{eqnarray}
with $\vec{q}=\vec{p}-\vec{p}'\;\;,\;\;\vec{k}=\frac{\vec{p}+\vec{p}'}{2}$
defined in terms of the in(out)going center of mass momenta
$\vec{p}(\vec{p}')$ of one nucleon.
Following eq.~(\ref{eq.nlo-lagrangean-3nf}), the three-body potential is given as
\begin{equation}\label{eq.pot_mom-3nf}
V^{(LO)}_{\textrm{EFT}_\slashed{\pi},3N}=\sum\limits_{i<j<k}^AC^{\textrm{LO}}_{3\textrm{NI}}\left(\vec{\tau}_i\cdot\vec{\tau}_j+
\vec{\tau}_k\cdot\vec{\tau}_i+\vec{\tau}_j\cdot\vec{\tau}_k\right)\;\;\;.
\end{equation}
Regularized with $f_\Lambda(\vec{q})=\text{exp}\left(-\vec{q}^2/\Lambda^2\right)$ and
Fourier transformed, eq.~(\ref{eq.pot_mom}) is cast into the form
\begin{small}
  \begin{eqnarray}\label{eq.pot-coord}
    V^{(NLO)}_{\textrm{EFT}_\slashed{\pi},NN}&=&
\sum\limits_{i<j}^AI_0\left(\Lambda,r\right)\left(A_1+A_2\vec{\sigma}_i\cdot\vec{\sigma}_j\right)+\left(A_3+A_4\vec{\sigma}_i\cdot\vec{\sigma}_j\right)\Big\lbrace I_0\left(\Lambda,r\right),\vec{\nabla}^2\Big\rbrace+\nonumber\\
&&I_0\left(\Lambda,r\right)\left(A_5+A_6\vec{\sigma}_i\cdot\vec{\sigma}_j\right)\vec{r}^2+\nonumber\\
    &&I_0\left(\Lambda,r\right)A_7\vec{L}\cdot\vec{S}+I_0\left(\Lambda,r\right)A_8\left(\vec{\sigma}_i\cdot\vec{r}\vec{\sigma}_j\cdot\vec{r}-
\frac{1}{3}\vec{r}^2\vec{\sigma}_i\cdot\vec{\sigma}_j\right)\nonumber\\
&&-A_9\Bigg\lbrace I_0\left(\Lambda,r\right),\Big[\big[\partial^r\otimes\partial^s\big]^{2}\otimes\big[\sigma_1^p\otimes\sigma_2^q\big]^{2}\Big]^{00}\Bigg\rbrace\;,
  \end{eqnarray}
\end{small}
and eq.~(\ref{eq.pot_mom-3nf}) into
\begin{equation}\label{eq.pot-coord-3nf}
V^{(LO)}_{\textrm{EFT}_\slashed{\pi},3N}=\sum\limits_{\stackrel{i<j<k}{\textrm{cyclic}}}^AI_0\left(\Lambda,r_{ij}\right)I_0\left(\Lambda,r_{jk}\right)C^{\textrm{LO}}_{3\textrm{NI}}\,\vec{\tau}_i\cdot\vec{\tau}_j\;\;\;,
\end{equation}
with the interparticle vector $\vec{r}=\vec{r}_i-\vec{r}_j$, the orbital angular
momentum operator ${\vec{L}=-i\vec{r}\times\vec{\nabla}}$,
and total spin operator $\vec{S}=\frac{1}{2}(\vec{\sigma}_i+\vec{\sigma}_j)$,
as well as the regulator function
$I_0\left(\Lambda,r\right)=\text{exp}\left(-\Lambda^2\vec{r}^2/4\right)$.
It is not necessary to symmetrize the potential with respect to particle exchange because the basis states are
antisymmetric.
Instead of fitting the original LECs $C_i$, it was favorable to adjust directly linear combinations
\begin{equation}\label{eq.lec-redef}
  A_i=\sum_{j=1}^9a_{ij}\Lambda^{n(j)}C_j\;\;\;.
\end{equation}
This re-definition avoids a fine-tuning of large versus small terms in the sum
which arises because the anticommutators absorb powers of $\Lambda$ into the
derivative acting on the regulator function $I_0\left(\Lambda,r\right)$.

The operator sets of AV18 and
CD-Bonn are subsets of the one of eq.~(\ref{eq.pot-coord}), while the
relatively complicated radial dependences of the former contrasts with the
polynomial-weighted Gaussians of eq.~(\ref{eq.pot-coord}).  Although most of
the operator structure is hidden in the anti-commutators, Hermitecity
is manifest in this form,
and the Gaussian radial dependences allow for efficient RRGM calculations.
The natural size of the coefficients $A_i$ is not trivial to estimate
because the derivatives in the two anticommutators act not only on the wave
function but also on the Gaussians in the potential.  Therefore, the
parameters $A_{3,4,9}$ are of lower power in $\Lambda$ and hence differ
considerably in size relative to the others.

The NN P-wave amplitudes are $\text{N}^3$LO in EFT$_\slashed{\pi}$.
Hence, four constraints,
\begin{equation}\label{eq.p-wave-hard}
\langle ^{2S+1}P_J\vert V^{(NLO)}_{\textrm{EFT}_\slashed{\pi},NN}\vert ^{2S+1}P_J\rangle=0\;\;\;,
\end{equation}
can be employed to reduce the number of parameters to five.
Here, we do not impose those constraints exactly, but instead extend the $\chi^2$-measure
to fit the P-wave phase shifts to a fraction of the Nijmegen values,
\begin{equation}\label{eq.p-wave-loose}
\delta_{\textrm{fit}}\left(^{2S+1}P_J\right)\leq 0.1\delta_{\textrm{Nij}}\left(^{2S+1}P_J\right)\;\;\;.
\end{equation}
This is fully consistent with the EFT philosophy that higher order
interactions can only induce higher order corrections in observables.
Different P-wave constraints, compatible with eq.~(\ref{eq.p-wave-loose}), allow for such a controlled modification
of short-distance structure.
An exact implementation of eq.~(\ref{eq.p-wave-hard}) provides no significant
gain in computer time relative to eq.~(\ref{eq.p-wave-loose}). We therefore exploit the
additional handle on higher order effects provided by eq.~(\ref{eq.p-wave-loose})
to gauge the accuracy of a NLO calculation. A consequence of this approach are non-vanishing
rank one and two interactions corresponding to LECs $A_7$ and $A_{8,9}$, respectively.
These tensor structures are also found when one implements the SD-interactions, which
enter beyond NLO. We choose to include these as representations of higher-order effects
and to constrain their parameters by the SD-mixing angle.

Different sets of weight factors for the phase- and $B(d)$-deviations were used for the various
potentials. Therefore, this $\chi^2$ is no objective criterion for the quality of
the potential and we abstain from quoting it.

The basis of this potential does not mix the singlet with the triplet NN channels.
Therefore, it was convenient to fit the projections of the potential in the
spin singlet- and triplet channels separately.
In the former, three parameters were adjusted to reproduce $\delta_{\textrm{Nij}}(^1S_0)$ and
$\delta_{\textrm{fit}}(^1P_1)$.  In the triplet channel, the fit is initially
to $B(d)$ only, followed by a fine-tuning of the six LECs to refine $\delta_{\textrm{fit}}(^3S_1)$,
$\delta_{\textrm{fit}}(^3P_{0,1,2})$ and the SD-mixing angle
$\epsilon_{1\,\textrm{fit}}(^3S_1-^3D_1)$.
Here, the following hierarchy of weights $w(\textrm{observable})$ in the
$\chi^2-$function was used: $w(B(d))>w\left(\delta(^{3,1}S_{1,0},\epsilon_1)\right)>w\left(\delta(^{2S+1}P_J)\right)$,
\textit{i.e.} the most weight was put onto $B(d)$, and we typically used a ratio of $w(B(d))/w(\delta)$ of the order of $10$.
Dependence of the resulting values
for the nine LECs on the model space was minimized by using an almost complete
set of 40 Gaussian basis states as described in sect.~\ref{ch.rrgm}.

The $\chi^2$-minimization which determined the LECs was carried out with a
modified version of the genetic search algorithm already used for wave
function optimizations in~\cite{genalg}. This algorithm has the advantage of
being independent from an educated guess for an initial set of parameters.
With the terminology specified in~\cite{genalg}, the search parameters for the
algorithm were chosen as follows. The initial population was set to consist of
more than $10^4$ individuals, with each individuum corresponding to a set of
LECs. This relatively large number should ensure that a good fraction of the
entire parameter space is probed. The search intervals used for the $A_i$'s
were chosen differently: $A_{3,4,9}\in[-10^3;+10^3]$, while
$A_{1,2,5-8}\in[-10^4;+10^4]$.
A factor of up to $\Lambda^4$ is not included in the $A_{3,4,9}$ but contributes
when the derivatives in the anticommutator act on the regulator functions (see
above). Therefore, we expect the $A_{3,4,9}$ to be smaller than the other $A_i$'s.

The contribution a specific operator made to $B(d)$ differed amongst the $^iV_\slashed{\pi}$.
While some EFT$_\slashed{\pi}$ potentials distribute $B(d)$ similar to the potential models,
AV18~\cite{av18} and CD-Bonn~\cite{bonn}, \textit{i.e.} the largest fractions coming
from the central, and tensor terms, others summed $B(d)$ quite differently.
However, there is no physical reason why one operator of $^iV_\slashed{\pi}$
should contribute more than the other.
Not only are LO and NLO operators combined in
$V^\textrm{NLO}_{\textrm{EFT}_\slashed{\pi}}(\vec{r})$ (see
eq.~(\ref{eq.pot-coord})), but they also have portions of the two central
operators hidden in the central anticommutator, as well as a tensor component
in the rank two anticommutator.

A heuristic explanation for the emergence of a variety of different LEC sets
each representing a valid EFT$_\slashed{\pi}$ interaction shall now be given.
The internal structure of the nucleons is encoded in the LECs, and different regulators
lead, after proper renormalization, to the same low-energy NN observables.  Therefore, no unique set
of coupling constants is expected.
Here, two methods are employed to model different short-distance interactions with the same long distance behavior. First, the regulator
was altered by changing the magnitude of the cutoff.
Second, different sets of LECs were found by using different input for the fit.

The difference in input data must be compatible with the order at which
the EFT calculation is carried out in the following sense. A consistent EFT
calculation at order $n$ predicts low-energy observables accurately up to
uncertainties of order $Q^{n+1}$ in the dimensionless expansion parameter. This
property extents to the data used to fit the LEC, \textit{i.e.} there is no need
to refine the fit of the parameters in table~\ref{tab.pots} for a better reproduction of, \textit{e.g.} the deuteron binding
energy $B(d)$ of the potentials. By weighting data differently in
the $\chi^2$-function, different sets of LECs were found, see table~\ref{tab.pots}. The values for $B(d)$
and the neutron-proton phase shifts for those sets were all in the NLO uncertainty
range around data, see table~\ref{tab.pot-nn-obs}.
Another approach which yields potentials with the same long-distance but different
short-distance structure is changing the input for the P-waves.
The P-wave phase shifts are observables of higher order, and therefore the values
predicted by the NLO potentials are only restricted to a range compatible with zero. We enforce different
P-wave interactions by including and varying an appropriate
term in the $\chi^2$-function, see eq.~(\ref{eq.p-wave-loose}). We employed a combination of both methods, different weights
and different P-wave input, to find several NN phase shift equivalent
potentials for a single cutoff value.

Another point had to be kept in mind for the fit. In principle, the EFT
philosophy allows for deeply bound states because reactions at
energies for which the pion-less theory is applicable will not probe those
low-lying, nonphysical states.
As mentioned in sect.~\ref{ch.piless}, they can occur when re-summing effective range contributions.
Nevertheless, we considered only potentials which were
found not to posses any of these deeper lying bound states. The reason is that
the variational basis does not span the entire Hilbert space and therefore
might be insufficient to expand a low-lying and hence very localized ground
state of a potential in the course of the fit.  In that case, it would be
impossible to tell if this state is low enough in energy to be considered
marginal. To ensure that no such ghost states were present during the fit,
vectors with narrower width parameters were added to the basis for the
deuteron and the triton to model more localized wave functions and sort out
the troublesome LEC parameter sets. Unphysical states might also be formed as
clusters in a three- or four-nucleon calculation because of narrower basis
states, in which case the corresponding LEC set was sorted out.
We employed as
figures of merit for a potential a converged value of the $\chi^2$-function
after successive runs of the genetic algorithm, and the numerical stability of
the deuteron and triton binding energies with respect to changes in the model
space, see discussion below.
\setlength{\tabcolsep}{0.2em}
\begin{table*}
  \caption{\label{tab.pots}Numerical values of the cutoff $\Lambda$ and the LECs of the NLO EFT$_\slashed{\pi}$ potentials.
    The LECs were fitted to $B^{\textrm{exp}}(d)$~\cite{exp-deut} and NN phase shifts.}
\footnotesize
    \begin{tabular}{ccccccccccc}
      &$\Lambda$&$A_1$&$A_2$&$A_3$&$A_4$&$A_5$&$A_6$&$A_7$&$A_8$&$A_9$\\
      &\small{[MeV]}&\small{[MeV]}&\small{[MeV]}&\small{[MeV$\cdot$fm$^2$]}&\small{[MeV$\cdot$fm$^2$]}&\small{[MeV$\cdot$fm$^{-2}$]}&
      \small{[MeV$\cdot$fm$^{-2}$]}&\small{[MeV]}&\small{[MeV$\cdot$fm$^{-2}$]}&\small{[MeV$\cdot$fm$^{2}$]}\\

\hline\hline
      $^1V_\slashed{\pi}$&414&-143&464&-57.0&68.8&-52.4&-346&-3.97&-62.1&-0.111\\
      $^2V_\slashed{\pi}$&432&-612&944&-157&110&351&-723&66.0&-168&-0.137\\
      $^3V_\slashed{\pi}$&544&-1224&1036&-432&336&1704&-1851&-78.9&-378&-0.130\\
      $^4V_\slashed{\pi}$&544&-1625&-89.6&-298&49.8&1870&34.3&-990&-734&$2.13\cdot 10^{-3}$\\
      $^5V_\slashed{\pi}$&544&-405&317&-125&108&287&-493&-271&-360&$-9.88\cdot 10^{-4}$\\
      $^6V_\slashed{\pi}$&648&84.9&-324&-888&404&4845&-2064&-342&-2319&-0.716\\
      $^7V_\slashed{\pi}$&648&-1316&1039&-530&431&3775&-3666&-75.4&-1052&-0.178\\
      $^8V_\slashed{\pi}$&672&301&-143&-708&480&3527&-3450&-192&-1613&-0.257\\
      $^9V_\slashed{\pi}$&672&-158&-201&-362&275&2001&-1659&79.7&-1270&-0.0850\\
\hline
    \end{tabular}
\end{table*}
\begin{table}
  \caption{\label{tab.pot-nn-obs}Two-nucleon observables calculated with NLO EFT$_\slashed{\pi}$ potentials as defined in eq.~(\ref{eq.pot-coord}) and table~\ref{tab.pots}.
    The deuteron binding energies $B(d)$ are labeled by the RRGM model space in which they
    were calculated. The data for the scattering lengths $a_{s,t}$ and the effective ranges $r_{s,t}$ is taken from~\cite{exp-anp}.}
\footnotesize
    \begin{tabular}{ccccccccc}
      &$\Lambda$&$B^{w_{120}}_d$&$B^{w_{12}}_d$&$B^{w_{63}}_d$&$a_s$&$a_t$&$r_s$&$r_t$\\
      &\rotatebox{90}{\tiny{[MeV]}}&\rotatebox{90}{\tiny{[MeV]}}&\rotatebox{90}{\tiny{[MeV]}}&
      \rotatebox{90}{\tiny{[MeV]}}&\rotatebox{90}{\tiny{[MeV$^{-1}$]}}&\rotatebox{90}{\tiny{[MeV$^{-1}$]}}&\rotatebox{90}{\tiny{[MeV$^{-1}$]}}&\rotatebox{90}{\tiny{[MeV$^{-1}$]}}\\
\hline\hline
      \begin{small}exp\end{small}&-&2.225&2.225&2.225&-0.120&0.0275&0.0138&0.00892\\ \hline
      $^1V_\slashed{\pi}$&414&2.239&2.238&2.226&-0.121&0.0275&0.0136&0.00914\\
      $^2V_\slashed{\pi}$&432&2.224&2.212&2.203&-0.119&0.0279&0.0134&0.0104\\
      $^3V_\slashed{\pi}$&544&2.221&2.205&2.153&-0.123&0.0268&0.0144&0.00824\\
      $^4V_\slashed{\pi}$&544&2.233&2.201&2.185&-0.119&0.0270&0.0133&0.00859\\
      $^5V_\slashed{\pi}$&544&2.224&2.222&2.186&-0.119&0.0258&0.0133&0.00666\\
      $^6V_\slashed{\pi}$&648&2.212&2.019&2.034&-0.120&0.0271&0.0132&0.00835\\
      $^7V_\slashed{\pi}$&648&2.214&2.137&2.012&-0.115&0.0260&0.0124&0.00687\\
      $^8V_\slashed{\pi}$&672&2.221&1.985&2.124&-0.118&0.0264&0.0125&0.0077\\
      $^9V_\slashed{\pi}$&672&2.250&2.182&2.114&-0.118&0.0255&0.0125&0.00648\\
\hline
    \end{tabular}
\end{table}
The numerical values of the LECs over a range of cutoffs are presented in table~\ref{tab.pots}.
The results for the deuteron binding energy in three
different model spaces per potential are listed in table~\ref{tab.pot-nn-obs} to demonstrate the
approximate closure of the set $w_{120}$ which was used to fit the LECs. A
model space dependence of the potential would lead to different deuteron binding energies $B(d)$ in
$w_{12}$ and $w_{120}$. Those sets were 40-dimensional, while the set
$w_{63}$ is only $9$-dimensional, with 6(3) optimized widths for the S(D) wave function component.
It was used in the $A=4$ calculations to build the deuteron fragments.
The change in $B(d)$ associated
with the substitution of $w_{120}$ by $w_{12}$ increases with the cutoff.
This is due to the absence of narrower width parameters in
$w_{12}$ which, as seen above, has a greater impact for narrower potentials, or larger
cutoffs, respectively.  Hence, this change is attributed to a
shortcoming of the $w_{12}$ model space rather than to one of the potential.
For the potential $^8V_\slashed{\pi}$, $B(d)$
decreases considerably from $w_{120}$ to $w_{12}$ but increases again and
approaches the $w_{120}$ value in a much smaller but optimized space
$w_{63}$. In general, the quality of the model space used for the fitting
procedure is comparable to the one used for the four-nucleon calculations presented
in the next section.

The decrease in $B(d)$ when going from $w_{120}$ to $w_{63}$ has no substantial
impact on the results for two reasons.  First, $w_{63}$ was only used to
expand the deuteron cluster in the \mbox{4-Helium} system and not the one in the
triton/\mbox{3-Helium}, which was built and optimized independently of the deuteron parameters.
Second, in the $A=4$ system only scattering reactions at energies below the
deuteron-deuteron threshold were considered. Therefore, the deuteron
configurations contribute only through the distortion channels.
This leads to the non-converged results for the imaginary part
of the 4-nucleon scattering length, see sect.~\ref{ch.res.a0}.

The width parameters which constitute set $w_{63}$ were tailor-made for
each potential using the genetic algorithm. In this process, the $w_{63}$
found for smaller cutoff values consisted on average of wider Gaussian width
parameters compared to the ones for relatively large values of $\Lambda$.
This behavior was already explained in sect.~\ref{ch.potfit} in the discussion
of fig.~\ref{fig.bd-cutoff}.
\begin{figure}
  \includegraphics[width=\columnwidth]{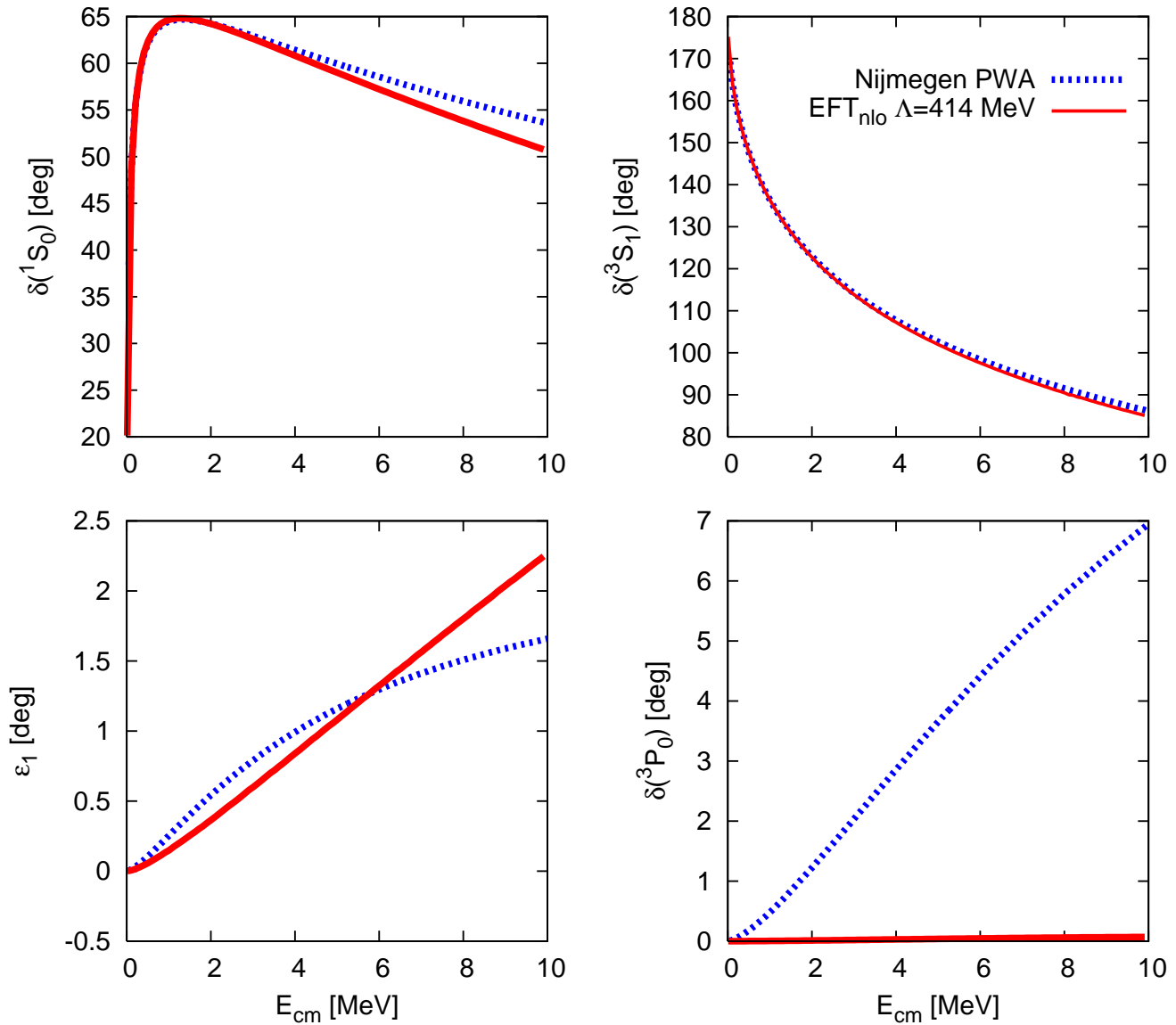}
  \caption{\label{fig.phases} Selected neutron-proton scattering phase shifts
    for the $^1V_\slashed{\pi}$ potential and values from the Nijmegen partial
    wave analysis~\cite{nijm,nnon} which were used to fit the LECs.}
\end{figure}
While $B(d)$ was used as experimental input to determine the LECs, the singlet
and triplet scattering lengths $a_{s,t}$ were used only implicitly through
their dependence on the phase shifts in the respective channels.  They were
calculated from the appropriate S-matrix elements at a center of mass energy,
$E_{\textrm{cm}}=1$~keV (see discussion below, eq.~(\ref{eq.scatt-length})).

As an example, the phase shifts for $^1V_\slashed{\pi}$ are shown in fig.~\ref{fig.phases}
along with the phases from the Nijmegen analysis~\cite{nijm,nnon}.
Only phase shifts for a center of mass energy below $0.35~$MeV were used to determine
the LECs, so that phases for higher energies can be used to gauge the quality of the fit.
The other potentials $^{2-9}V_\slashed{\pi}$ reproduce the Nijmegen phases with the same
accuracy, \textit{i.e.}, the phases up to $10~$MeV differ by less than $1$\%.
While $\delta_{\textrm{Nij}}(^1S_0)$ and $\delta_{\textrm{Nij}}(^3S_1)$ are
reproduced up to $1$\% for $E_{\textrm{cm}}\lesssim 10~$MeV, the SD-mixing
angle $\epsilon_1$ reaches, in the same energy interval, only a $10$\%
accuracy and also showed a stronger variation between the various potentials.
Both uncertainties are within the error margins predicted by
EFT$_\slashed{\pi}$, \textit{i.e.}, approximately $10$\% for the S-waves at typical
momenta $p_\text{typ}\approx 45~$MeV and $30$\% for the higher-order observable $\epsilon_1$.
For all potentials $^iV_\slashed{\pi}$ the
results for the phase shifts in the $^3D_1$ channel were close to zero,
$\delta(^3D_1)\approx 0$.  The $^3P_0$ channel is displayed to exemplify that
the P-wave interactions of the potentials are indeed small.

Finally, observe in table~\ref{tab.pot-nn-obs} that our cutoff values do not constrain
the effective range to unphysical values, as alluded to in the discussion of the Wigner
bound in sect.~\ref{ch.piless}.

Table~\ref{tab.pots} and fig.~\ref{fig.phases} demonstrate: (i) The quality of
the fit is sufficient for a NLO calculation with the expected accuracy level
of about $10$\% for EFT$_\slashed{\pi}$ at NLO. (ii) The potentials in the
2-nucleon sector are approximately equivalent.

\section{Results for three and four nucleons}\label{ch.res}

We now present the results of our feasibility study of using the RRGM with EFTs,
exemplified at NLO calculations in EFT$_\slashed{\pi}$. The section is divided
into four subsections. The first reports on an analysis of the correlation between
the triton charge radius and its binding energy. The second looks at the
splitting of the trinucleon binding energies due to Coulomb effects.
The third discusses the Tjon correlation, and the fourth contains the findings
for the \mbox{3-Helium-neutron} scattering system.

\subsection{Three nucleons: triton charge radius}\label{ch.res.friar-line}
\begin{figure}
  \includegraphics[width=\columnwidth]{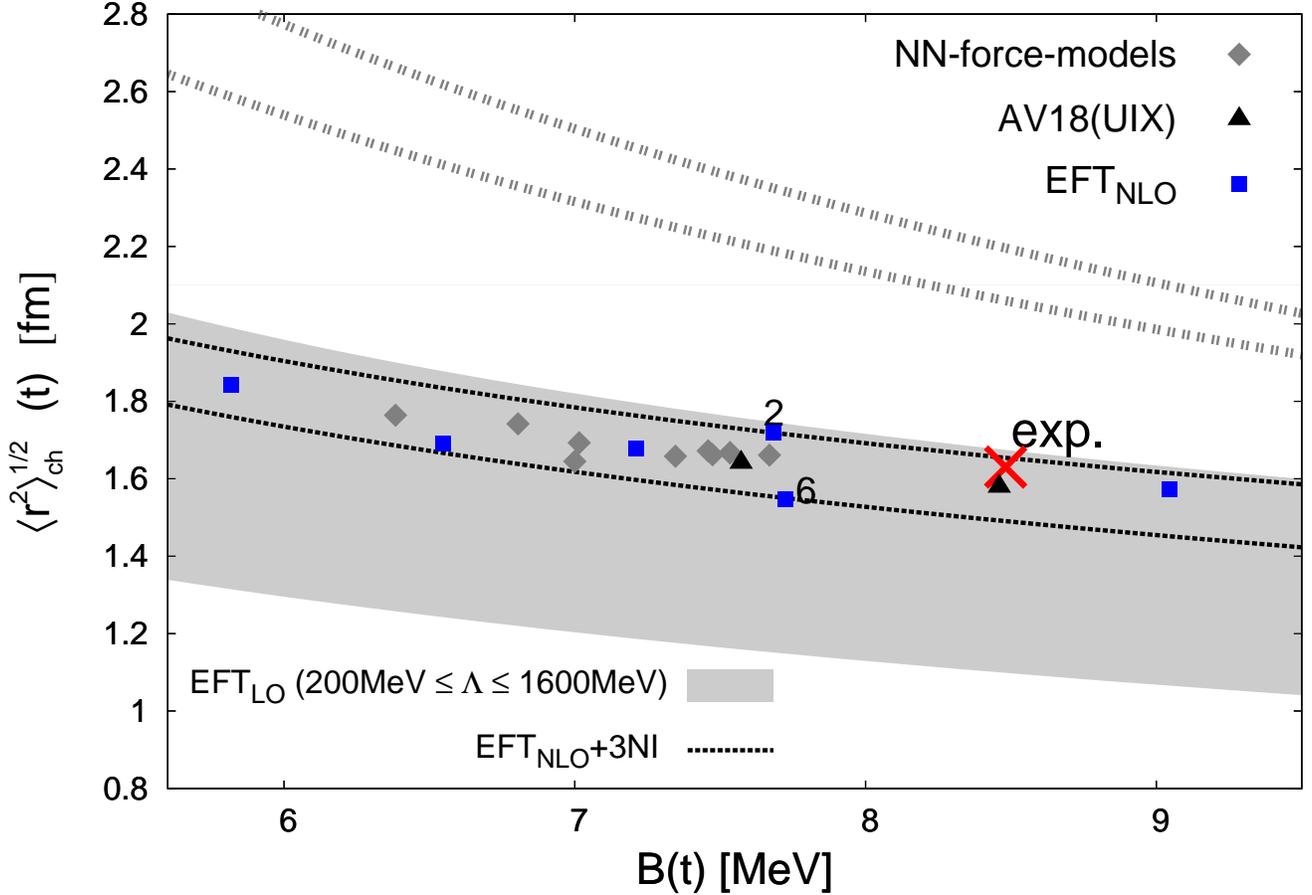}
  \caption{\label{fig.rch} The correlation between the triton charge radius
    and its binding energy.
The narrower band mapped out by the EFT$_\slashed{\pi}$ NLO results from the RRGM (blue squares, black dashed lines) as described in the text,
compared to our RRGM LO results (gray shaded area) and a LO calculation by~\cite{platter-rch} (dashed gray lines), demonstrates convergence,
also to the datum ($r_\textrm{ch}^t$ from~\cite{exp-t-rch}, $B^\textrm{exp}(t)$ from~\cite{exp-bt}). The values from AV18(+UIX) (black triangles, RRGM
calculation for this work) and a variety of other potential models (gray diamonds~\cite{rch-pots-1,rch-pots-2}) are consistent with
    our proposed EFT$_\slashed{\pi}$ NLO correlation band.
}
\end{figure}
Friar \textit{et al.}~\cite{friar-rch} observed a correlation between the
triton binding energy $B(t)$ and the triton charge radius using various
nuclear force models, before Platter \textit{et al.}~\cite{platter-rch} showed by a variation of
the three-body interaction parameter at leading-order that it can be understood as a
consequence of universality in EFT$_\slashed{\pi}$. With the model spaces
defined in sect.~\ref{ch.rrgm}, $B(t)$ and the corresponding ground state wave
function $\langle\vec{r}|t\rangle$ were calculated at LO and NLO, and from that the root mean square charge radius
$r^t_{\textrm{ch}}$ of the triton:
\begin{equation}\label{eq.rch-t}
r^t_{\textrm{ch}}=\left(\langle t|\sum_{i=1}^3\frac{1}{2}\vec{r}^2_i\left(1+\tau^3_{i}\right)|t\rangle\right)^{\frac{1}{2}}\;\;\;,
\end{equation}
where $\vec{r}_i$ is the position and $(1+\tau^3_{i})$ the charge
operator of the $i$-th nucleon.
The results in fig.~\ref{fig.rch} confirm the expected behavior of an increasing
$r^t_{\textrm{ch}}$ for more loosely bound systems.
Results of two EFT$_\slashed{\pi}$ LO calculations are shown.
The RRGM LO potentials map out a band (gray shaded area) which includes the datum. The lower and upper LO line
from~\cite{platter-rch}  (dashed gray lines in fig.~\ref{fig.rch}) result from a fit of the LO LECs $C_{1,2}^\textrm{LO}$
to either $a_{s},a_{s}$ or $B(d),a_s$, respectively, providing some measure of higher order effects. For the RRGM LO calculation, we used $a_{s,t}$ to
determine the LECs, and varied the cutoff from $200~\textrm{MeV}$ (top edge) to $1.6~\textrm{GeV}$ (lower edge)
in steps of $100~\textrm{MeV}$.
We choose the lower bound for $\Lambda$ to be in the region of the pion mass. The upper
bound is set by the observation that there is essentially no change for
cutoffs larger than $1.6~$GeV. The band appears to be saturated, with the bulk
of its width coming from the region $\Lambda\in[400;800]~$MeV.  For fixed
$\Lambda$, the 3NI was finally tuned to a given $B(t)$ to generate the
correlation lines.

The quasi-exact Faddeev calculation of~\cite{platter-rch} and the RRGM
results do not overlap. Since app.~\ref{ch.rrgm.numstab} demonstrates that numerical inaccuracies
of the RRGM are negligible, we speculate that this can be traced to the
differences between the regularization schemes employed. The Faddeev
calculation uses the separable cutoff function $f_\kappa(\vec{p},\vec{p}')=\text{exp}\left(-\vec{p}^2/\kappa^2\right)\text{exp}\left(-\vec{p}'^2/\kappa^2\right)$
with $\kappa\geq 1600~\textrm{MeV}$, while the RRGM uses
the non-separable regulator $f_\Lambda(\vec{p},\vec{p}')=\text{exp}\left(-\left(\vec{p}-\vec{p}'\right)^2/\Lambda^2\right)$,
plus an implicit regulator imposed by
the finite number of width parameters. Therefore, both methods combined can be
viewed as providing a check of residual regularization-scheme dependence.  A
conservative estimate of LO effects is thus the range of results coated by the
combination of both methods. The LO accuracy at the physical triton binding
energy is thus $\pm0.6~\textrm{fm}$, and the measured charge radius happens to lie
right in the middle of the LO band. Additional regularization schemes,
\textit{e.g.}, with a range of cutoff values in the Faddeev approach, are thus
conjectured to lead to an overlap between the results of~\cite{platter-rch}
and the RRGM bands.

At NLO, the potentials without 3NI map out a correlation band which is more
narrow than its LO counterpart, and again contains the datum. The width of the
band can be estimated by considering the difference between the results using
the extreme cases provided by potentials $^2V_\slashed{\pi}$ and $^6V_\slashed{\pi}$.
They produce the same NN scattering lengths and nearly the same triton binding energy, and their deuteron
binding energies differ by less than $10$\%, while their triton charge radii differ by
about $10$\%, consistent with the expectation of a NLO calculation. Finally, they
are based on two significantly different cutoff values, $\Lambda(^2V_\slashed{\pi})\approx
400~$MeV and $\Lambda(^6V_\slashed{\pi})\approx 650~$MeV. We therefore can base an estimate of
the NLO band on the range mapped out by varying the 3NI for these two
potentials. These lines are included in fig.~\ref{fig.rch}.

Variation of the cutoff, of the fitting-input, and of the 3NI lead therefore all
to similar assessments of the uncertainty of the theory at NLO. At fixed $B(t)$,
the charge radius varies from LO to NLO by $\lesssim 30$\%, in agreement with the
power-counting which estimates the correction from order to order to scale as ${Q\sim p_\textrm{typ}/{\Lambda_b}\approx \frac{1}{3}}$.

All three values, namely, $\pm 0.2~$fm from the na\"ive estimate $Q\approx\frac{1}{3}$
and the observed convergence from LO to NLO, and $\pm 0.1~$fm from the above mentioned
difference between $^2V_\slashed{\pi}$ and $^6V_\slashed{\pi}$, would be equally valid
estimates for the theoretical uncertainty of this NLO calculation.
As demonstrated in app.~\ref{ch.rrgm.numstab}, errors induced by numerical
inaccuracies are negligible.
Using the more conservative error estimate and the assumption for the NLO error band center
at $1.6~$fm at the experimental $B(t)$, EFT$_\slashed{\pi}$ predicts a value 
\begin{equation}\label{eq.nlo-rch}
r_{ch}^{t,\textrm{NLO}}=(1.6\pm 0.2)~\textrm{fm}
\end{equation}
within the band of the leading-order value $r_{ch}^{t,\textrm{LO}}=(2.1\pm 0.6)~$fm as quoted from~\cite{platter-rch}.
The NLO value is found in good agreement with experiment~\cite{exp-t-rch}, $r_{ch}^{\textrm{exp}}=(1.63\pm 0.03)~$fm.

Another argument in favor of our definition of the correlation band is provided by the
results of the two phenomenological models, AV18~\cite{av18} and AV18+UIX~\cite{uix}. In general, EFT predicts that
the results of a potential which reproduces or shares input observables at least to the accuracy
required at the considered order, deviate from the results of an appropriate EFT
potential by less than the theoretical uncertainty of the EFT values at this order in the
applicability range of the EFT.
This criterion is easily met by AV18(+UIX), and hence its predictions have to be
consistent with the proposed correlation band. The two-body potential AV18, reproducing the
Nijmegen phases much more accurately than required to fall into this category of potentials,
is expected to yield a value at a position within the $(r_{ch}^{t}-B(t))$-band. The prediction
for the triton charge radius of AV18+UIX, however, is expected to deviate less than $10$\% from
the experimental datum within the error band, because this model has a three-body interaction added to reproduce the experimental $B(t)$.
Both expectations are consistent with the results shown in fig.~\ref{fig.rch}.

In conclusion, the results for these 3-nucleon observables show that although the
potentials are approximately NN phase shift equivalent, they differ
in their predictions of three-body observables. As mentioned in the introduction,
a proper renormalization of the theory requires therefore one 3-nucleon contact interaction.
Setting this 3NI to zero in the potentials $^{1-9}V_\slashed{\pi}$,
the expected dependency  of observables in $A>2$ systems on how the unobservable short-distance physics is modeled is observed.
Different short-distance physics is modeled by the potentials not only by varying cutoff values but also by
differing sets of LECs for the same cutoff, while the scheme- and
regulator-dependent three-body interaction parameter is chosen as
zero. With this parameter fitted to the triton binding energy, the prediction for the triton charge radius is consistent
with experiment within the expected uncertainty range. A significant convergence from LO to NLO is observed.
\subsection{Three nucleons: effect of the Coulomb force}\label{ch.res.t-3he}
\begin{figure}
  \includegraphics[width=\columnwidth]{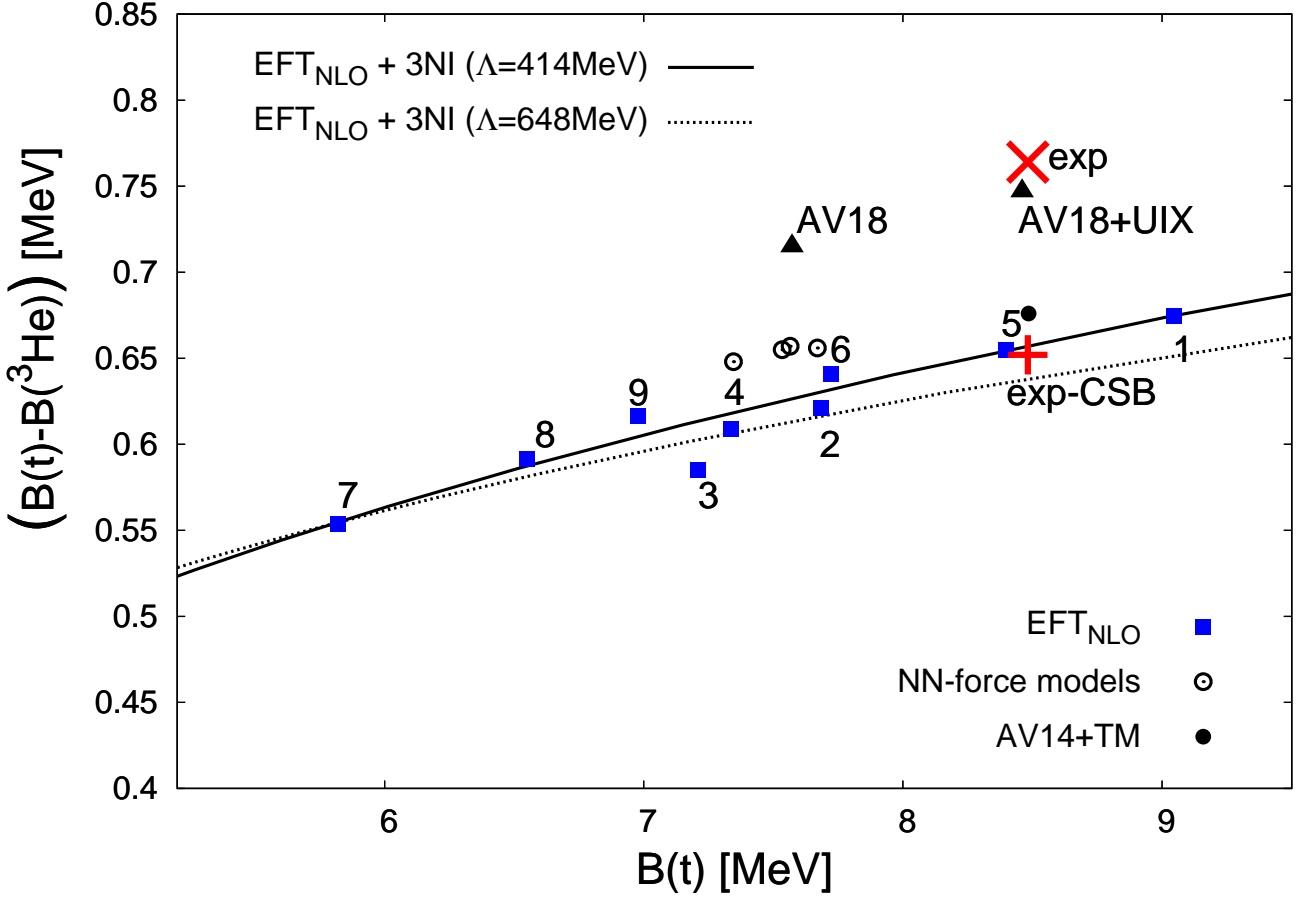}
\caption{\label{fig.trinucl-splitt} Binding energy difference between $^3$He
  and $^3$H for EFT$_\slashed{\pi}$ potentials (filled squares) compared to
  various isospin invariant potential models
  calculations~\cite{rch-pots-2,pest-pot} (circles) and RRGM values for
  AV18(+UIX) (triangles) which contain charge symmetry breaking terms.
  For the potentials $^{1}V_\slashed{\pi}$ (solid) and $^{7}V_\slashed{\pi}$ (dashed),
  a smooth variation of the 3NI leads to the two correlation lines.
  The upright cross is the experimental value without the contribution from CSB
  terms as found in~ \cite{csb-tni}.}
\end{figure}
In fig.~\ref{fig.trinucl-splitt}, the RRGM results for the splitting between
the binding energies of \mbox{3-Helium} and the triton are shown for the potentials
$^{1-9}V_\slashed{\pi}$. At NLO in the pion-less EFT, the strong interaction
is isoscalar and hence does not break charge symmetry.  Therefore, charge
symmetry breaking (CSB) comes in our RRGM calculation only from including the
Coulomb interaction between the protons in $^3$He. The EFT$_\slashed{\pi}$
results, hence, only show the model independent contribution of Coulomb interactions
to the trinucleon binding energy splitting.
The correlation band is mapped out by the EFT$_\slashed{\pi}$ potentials
with zero 3NI and by smoothly varying this
three-body parameter for potentials
$^{1,7}V_\slashed{\pi}$. Both approaches result in correlations consistent with each other
and with the results of the other NLO potentials.
At the experimental triton binding energy, this leads to predicting
\begin{equation}\label{eq.nlo-splitt}
  \left(B(t)-B({^3\textrm{\small He}})\right)=\left( 0.66\pm0.03\right)~\textrm{MeV}\;\;\;.
\end{equation} 
Here, we estimate the theoretical uncertainty by the spread of the
phase-equivalent NN potentials as in the previous section, e.g.~comparing
$^{3,9}V_\slashed{\pi}$. The \emph{a priori} error estimate at NLO of
$\lesssim10$\% gives a larger uncertainty of $\pm0.07~\textrm{MeV}$. However,
one should keep in mind that including iso-scalar strong interactions by
higher order terms of the effective-range expansion has identical effects on
the strong interactions inside the triton and ${}^3$He. These effects cancel
out in the difference and only survive indirectly, as the strength of the
Coulomb interaction in a system is also correlated to its size. We therefore
quote the width of the correlation band as error estimate of our calculation.

This value deviates by about $0.1~$MeV from the experimental value of
$0.764~$MeV, see \textit{e.g.} the recent review on CSB and Charge
Independence Breaking (CIB)~\cite{opper-csb}. In line with the argumentation
above, we attribute this difference to isospin breaking CIB/CSB interactions
coming from the explicitly broken chiral symmetry in the strong and
electro-weak sector from which only the parts resulting in the Coulomb force
have been considered in this calculation. They enter in EFT$_\slashed{\pi}$
only at higher order. To support this assertion, results of the potential
models AV18(+UIX), which contain CSB interactions, are included in
fig.~\ref{fig.trinucl-splitt}.  Both potentials are not elements of the
correlation band suggested by the EFT$_\slashed{\pi}$ points but agree with a
shifted band, centered around the datum. In contrast, the values from charge
symmetric potential models in fig.~\ref{fig.trinucl-splitt} lie within the NLO
EFT$_\slashed{\pi}$ band. This leads us to predict a model independent CSB/CIB
contribution to the binding difference in NLO EFT$_\slashed{\pi}$ at the
experimental triton binding energy of
\begin{equation}\label{eq.nlo-CSB}
  \left(B(t)-B({^3\textrm{\small He}})\right)^\text{CSB/CIB}=\left( 0.10\mp0.03\right)~\textrm{MeV}\;\;\;,
\end{equation}
anti-correlated with the Coulomb contribution to give the experimentally
established difference. This is to be compared with the contributions from
2- and 3-nucleon CSB interactions which stem from Chiral Effective Field
Theory, Breit and vacuum polarization corrections, and from corrections to the
kinetic energy operator. In Ref.~\cite{csb-tni}, these were calculated to sum
up to $(0.112\mp0.022)~$MeV, leaving about $(0.652\pm0.022)~$MeV for the soft
photon effects, dominated by the Coulomb interaction. This is in perfect
agreement with the EFT$_\slashed{\pi}$ result.
\subsection{Four nucleons: bound state}\label{ch.res.tjon}
Venturing into the four-body system, we now present results of the Tjon correlation line\cite{tjon}
including Coulomb interactions between the ground state energies of the triton
and \mbox{4-Helium} at LO and NLO, extending the LO analysis of\cite{platter-tjon}, where the effect of Coulomb
interactions was only estimated.

Before reporting the results, we describe the employed variational space and its construction.
The model space in which the \mbox{4-Helium} binding energy $B(\alpha)$ was initially calculated
is spanned by vectors defined in eq.~(\ref{eq.bsbv}) with relative angular
momenta $l_{ki}\leq 2$ including all coupling schemes to yield a total
angular momentum state $J^\pi=0^+$ and width parameters $\gamma_{dk}$ to
allow for the formation of triton, \mbox{3-Helium}, and deuteron fragments.  As
mentioned above, the $w_{63}$ sets were used for the deuteron, resulting in
169 configurations.  For the 3-nucleon fragments, a small model space of
dimension $d\leq 70$ was generated for each potential as follows. To an
initial set of 20 basis vectors which bind the triton, one new basis vector
was added, and its two width parameters $\gamma_{dk}$ (see
eq.~(\ref{eq.bsbv})) optimized with the genetic algorithm to maximize the gain
in binding energy. This process was iterated until
$|B^{224}(t)-B^{\textrm{small}}(t)|\leq 500~$keV, where the superscript labels
the aforementioned 224-dimensional model space. For \mbox{3-Helium}, a copy of this
model space was used which differed only in its isospin quantum numbers.
On average, 300 configurations were considered for the model space of
a given $V_\slashed{\pi}$, for each of which four or five inter-fragment width
parameters were used, resulting in roughly 1200 basis vectors.

\begin{figure}
  \includegraphics[width=\columnwidth]{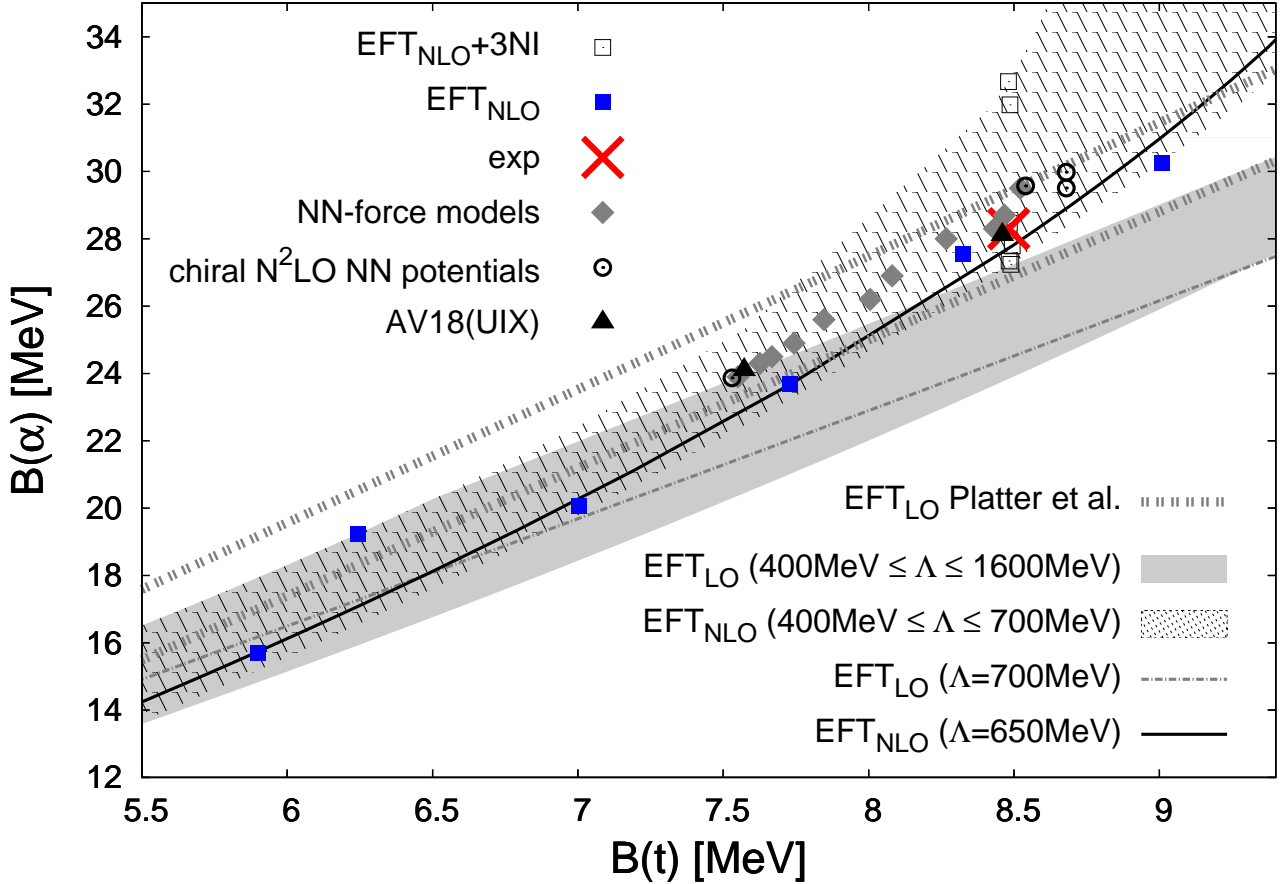}
  \caption{\label{fig.tjon} Correlation between the triton and $^4$He
    binding energies (Tjon line). The filled squares are the results using NLO
    EFT$_\slashed{\pi}$ NN potentials from table~\ref{tab.pots} with no 3NI, and the empty squares represent
    the predictions for the 3NI fitted to $B(t)$. The shaded NLO band results from
    a variation of the 3NI for those potentials.
    We compare to LO results from~\cite{platter-tjon}, where the upper (lower) dashed line was calculated with LEC fitted to $a_{s,t}$ ($B(d),a_s$), and to our
    RRGM calculation for $\Lambda$ ranging from $400~\textrm{MeV}$ (lower edge) to $1.6~\textrm{GeV}$ (top edge of light gray area).
    $B^\textrm{exp}(\alpha)$ is taken from~\cite{exp-ba}, the values for
    AV18(+UIX) are reported in~\cite{mythesis}, and those using a chiral expansion to NLO and N$^2$LO
    in~\cite{xpt-epelbaum}.}
\end{figure}
To assess if the reduction in the number of triton components
still guaranteed for an almost complete \mbox{4-Helium} model space, the binding
energy $B(\alpha)$ thus obtained was compared to one calculated in the much
larger scattering model space.
Nevertheless, the two values
differed by less than 50~keV. This model space, used for the $0^+-$channel
scattering calculation is defined below and is spanned by more than 7000 basis vectors.
This study of reducing the dimension of the 3-nucleon fragments in the $\alpha$-particle
without deviating significantly from the assumed converged value is crucial for
future applications of the method to $A>4$ systems.

In fig.~\ref{fig.tjon}, the results are compared to the LO band calculated by
Platter \textit{et al.}~\cite{platter-tjon}. The spread of the NLO values is not in conflict
with EFT$_\slashed{\pi}$ which allows a $\lesssim 10$\% uncertainty at NLO.
As for the triton charge radius, Platter  \textit{et al.} obtained the upper (lower) boundary of the LO band by choosing different NN
observables, $a_{s,t}(B(d),a_s)$, to fit the LO LECs
for cutoffs high enough so that $B(\alpha)$ did not change when further changing the cutoff.
They already pointed out that this gives only a
crude estimate of higher order effects. Figures \ref{fig.rch}
and~\ref{fig.tjon} confirm that the LO accuracy would be overestimated with this
method.
Our LO calculation uses the same interactions as in sect.~\ref{ch.res.friar-line}, with $a_{s,t}$ as
input, and cutoff values from $400~\textrm{MeV}$ (lower bound of LO band in fig.~\ref{fig.tjon}) to $1.6~\textrm{GeV}$ (upper bound).
In contrast to our RRGM LO correlation between $B(t)$ and the triton charge radius, the width of the RRGM LO Tjon band
is however not converged. The shifts between the positions of the correlation lines corresponding to cutoffs of $400~\textrm{MeV}$ (lower edge of
gray area in fig.~\ref{fig.tjon}), $700~\textrm{MeV}$ (dashed dotted line), and $1.6~\textrm{GeV}$ indicate that a variation
of $\Lambda$ even beyond $1.6~\textrm{GeV}$ would be necessary to assess the LO uncertainty
from cutoff variations only.
However, elaborate technical modifications are required for
the RRGM calculation at higher cutoffs. Analogous to the conservative estimate for the LO uncertainty in fig.~\ref{fig.rch},
the LO Tjon correlation band is thus mapped out by both the RRGM and the Faddeev/Yakubovsky
(thick dashed lines) results, which overlap nicely in this plot. This combined correlation band includes the datum
and the narrower NLO band, which, from fig.~\ref{fig.tjon}, has at the experimental $B(t)$ a width of about $5~$MeV centered around $28~$MeV and
results in a prediction of
\begin{equation}\label{eq.b4he-nlo}
B^{\textrm{NLO}}(\alpha)=(28\pm 2.5)~\textrm{MeV}\;\;\;,
\end{equation}
which is consistent with the expected NLO uncertainty of about $10$\% and with experiment.
Again, the results of the AV18(+UIX) models lie within the proposed band as it is expected
of all interaction models of at least NLO. The observed broadening of the correlation band
is a manifestation of the momentum dependence of the EFT expansion. The accuracy decreases
with increasing typical momentum, eventually leading to a breakdown of the expansion.

From the fact that there is still a one-parameter correlation, we conclude that no
four-nucleon contact interaction is required to renormalize the theory at
NLO. One three-body parameter fitted to data suffices to yield
proper NLO predictions for four-body observables within the theoretical
accuracy (empty squares in fig.~\ref{fig.tjon}).
\subsection{Four nucleons: scattering}\label{ch.res.a0}
\begin{figure}
  \includegraphics[width=\columnwidth]{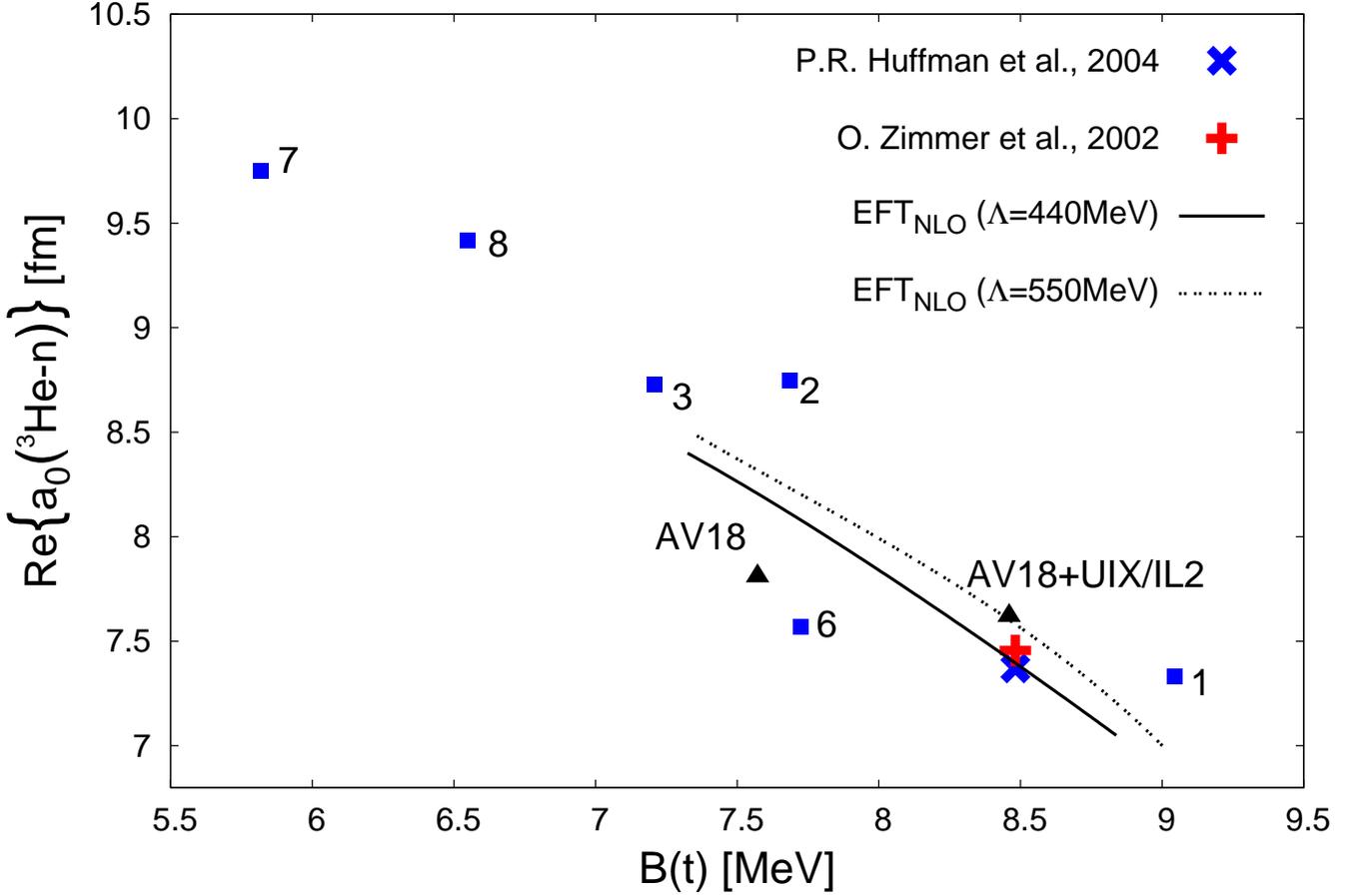}
  \caption{\label{fig.a0} The correlation between the triton binding energy
    and the real part of the spin singlet, S-wave scattering length
    $a_0(^3\textrm{{\small He-n}})$ for elastic $^3$He-n scattering. The dots
    represent the values of the NLO EFT$_\slashed{\pi}$ potentials with a
    numerical uncertainty of $\mathcal{O}(0.02~\textrm{fm})$. A variation of the 3NI
    yields the continuous correlation lines for two selected NLO potentials with a fixed
    cutoff.
}
\end{figure}

We now turn to scattering observables. In principle, all
low-energy observables should be correlated with the triton
binding energy. The recent results~\cite{delt-port-n3h} for
the singlet and triplet $n-^3$H scattering lengths using
three potential models and a N$^3$LO chiral potential are
evidence for this assertion in the four-nucleon scattering
system. Here, the real part of the S-wave spin singlet
scattering length $a_0(^3\textrm{{\small He-n}})$ for elastic
\mbox{3-Helium-neutron} scattering is investigated. In fig.~\ref{fig.a0},
its value is shown as a function of $B(t)$ for six potentials
$V_\slashed{\pi}$ including the Coulomb interaction.
For two NLO potentials with $\Lambda=440~\textrm{MeV}$ (solid line)
and $\Lambda=550~\textrm{MeV}$(dashed line), with 2-nucleon LECs fixed,
we also show the effect of a smooth variation of the 3-nucleon interaction parameter.

To extract $a_0(^3\textrm{{\small He-n}})$, six two-fragment channels,
$^3$He-n, t-p, d-d ($l_\textrm{rel}=0,2$), (nn)-(pp), and dq-dq (dq is the
singlet ``deuteron'' with $S=0$), are included.  The latter two consist of
unbound fragments and usually model possible three- and four-body breakup
reactions.  They, as well as the two d-d channels, are however for this
calculation only needed to provide configurations for distortion channels
since only the t-p channel is open a few eV above the $^3$He-n threshold.  For
the fragment wave functions $\psi_j$ (see eq.~(\ref{eq.ssbv})), a 224 dimensional basis
was used for the triton and \mbox{3-Helium}, and a 9 dimensional one for the
deuteron, whose six $L_j=0$ vectors built the nn, pp, and dq states.  For
these six channels, the 20 width parameters $w_{12}$ were used for the
$\omega_{jm}$ in eq.~(\ref{eq.ssbv}).  Almost all configurations included to
build those physical channels could be recycled as distortion channels to
allow for more freedom in the minimization of the variational functional.
Less than ten configurations had to be excluded to avoid numerical linear
dependences.  In each distortion channel, four to six relative width
parameters $\omega_{jm}$, taken from $w_{12}$ with
$\omega_{jm}>0.02~\textrm{fm}^{-2}$, were used.  Numerical stability and
convergence of $a_0(^3\textrm{{\small He-n}})$ were assessed by increasing the
number of included relative widths $\omega_{jm}$ by one for each distortion
channel, yielding changes in $a_0(^3\textrm{{\small He-n}})$ of the order of
the numerical uncertainties, given that the initial $\omega_{jm}$ were chosen
appropriately.  The lowest eigenvalue of the Hamiltonian is equal to the ground state energy of \mbox{4-Helium}
in this model space and was allowed to change in this process by not more than 10~keV.
Significantly larger changes in this eigenvalue which leads to a result not of
the order of magnitude suggested by the LO Tjon band signal numerical
linear dependences. If the model space is too small, or if the width parameters for
the relative wave function were chosen inappropriately, changes of the order of
$100~$keV up to a few MeV are expected.

The \mbox{3-Helium-neutron} scattering length $a_0(^3\textrm{{\small He-n}})$ was
calculated from the corresponding diagonal element $S_{22}$ of the S-matrix at
a center of mass energy of $E_\textrm{cm}<10~$eV above the $^3$He-n threshold
analogously to~\cite{hmh-4he}:
\begin{equation}\label{eq.scatt-length}
  a_0(E_\textrm{cm})=\frac{1-S_{22}(E_\textrm{cm})}{i\sqrt{\frac{3}{4}ME_\textrm{cm}}\big(1+S_{22}(E_\textrm{cm})\big)}\;\;\;.
\end{equation}
A fraction of the flux is diverted into the open triton-proton channel,
resulting in a nonzero $S_{12}$ S-matrix element and hence a nonzero imaginary
part of $a_0(^3\textrm{{\small He-n}})$ which is not reported here.
The 3NI variation yields an almost linearly increasing imaginary part with increasing $B(t)$ from about
$\textrm{Im}\lbrace a_0(^3\textrm{{\small He-n}})\rbrace\approx -7.5~\textrm{fm}$ at
$B(t)\approx 5.5~\textrm{MeV}$ to $\textrm{Im}\lbrace a_0(^3\textrm{{\small He-n}})\rbrace\approx -2.0~\textrm{fm}$
at $B(t)\approx 9.1~\textrm{MeV}$. This qualitative observation is consistent with the decreasing trinucleon
binding energy splitting with decreasing triton binding energy (see sect.~\ref{ch.res.t-3he}), which results in a
smaller separation between the respective thresholds in four-nucleon scattering.
This handle on the threshold separation suggests an approach to circumvent numerical
problems associated with the proximity of thresholds by extrapolating results for their physical
values from calculations performed at more deeply bound 3-nucleon states.

In fig.~\ref{fig.a0}, the predictions of the NLO potentials with and without 3NI
for $\textrm{Re}\lbrace a_0(^3\textrm{{\small He-n}})\rbrace$ decrease with increasing
triton binding energy. They map out a band which includes the datum.  A comparison
to hard sphere scattering qualitatively explains this behavior. A higher \mbox{3-Helium} binding
energy $B({^3\textrm{\small He}})$ corresponds to a smaller nucleus analogous to the
triton as shown in fig.~\ref{fig.rch}.  As the scattering length is
proportional to the radius of the hard sphere, $a_0(^3\textrm{{\small He-n}})$
is expected to decrease for increasing $B({^3\textrm{\small He}})$.
We define the NLO correlation band to be centered around the values predicted
by the potentials with zero 3NI.
The change in $a_0(^3\textrm{{\small He-n}})$ observed between $^2V_\slashed{\pi}$
and $^6V_\slashed{\pi}$ serves as an error estimate for the band centered around
$7.5~$fm. Again, the assumption of the center of the band at the experimental $B(t)$
follows from the values in fig.~\ref{fig.a0}.
Thus, EFT$_\slashed{\pi}^\textrm{NLO}$ reports
\begin{equation}\label{eq.a0-value-nlo}
\textrm{Re}\lbrace a_0(^3\textrm{{\small He-n}})\rbrace=\left(7.5\pm 0.6\right)~\textrm{fm}\;\;\;.
\end{equation}
The 3NI was varied only over the depicted range of trition binding
energies, where one and the same fixed RRGM variational space can be used.
The fragment model space for the two interactions was optimized
for the 3NI parameter fitted to $B(t)$. Only the number of included distortion channels
was changed to reach convergence within this space for each value of the 3NI.
The band mapped out by the two 3NI lines for
$\left(7.5~\textrm{MeV}~\lesssim B(t)\lesssim 9~\textrm{MeV}\right)$ includes the datum, and its slope is
consistent with the one indicated by the 6 NLO potentials without 3NI.
This explicitly demonstrates that a variation of the 3NI has the same effect as
varying the short-distance part of the NN interaction.
The NN model AV18 yields a value within the error band. Adding the UIX three-body
interaction moves this point into the $10$\% NLO uncertainty radius around the datum.
The numerical accuracy of this scattering calculation is assessed in app.~\ref{ch.rrgm.numstab-2} to be
better than $1$\%.

The conclusion drawn from the newly found correlation in fig.~\ref{fig.a0} is
that every potential with the correct NN low-energy phase shifts and
appropriately tuned three-body interaction, \textit{e.g.}, to give the correct
triton binding energy, predicts not only the correct $B(\alpha)$ but also
the experimental $a_0(^3\textrm{{\small He-n}})$ within a NLO error range.
In table~\ref{tab.a0}, the results for $a_0(^3\textrm{{\small He-n}})$ of RRGM
calculations with the EFT$_\slashed{\pi}^\textrm{NLO}$ potential and the AV18 NN force model with
and without the three-body interaction models Urbana9 (UIX) and Illinois2 (IL2)~\cite{ill}
are given.

\begin{table}
  \caption{\label{tab.a0} RRGM predictions for the triton binding energy and the spin singlet, S-wave scattering length $a_0(^3\textrm{{\small He-n}})$
    for elastic $^3$He-n scattering of phenomenological nucleon potential models. The values for AV18 and AV18+UIX are taken from~\cite{hmh-4he},
    while the AV18+IL2 numbers are new results of this work. The imaginary part was investigated on a qualitative level, only.}
    \begin{tabular}{cccc}
      force&$B(t)~$[MeV]&$\textrm{Re}\lbrace a_0\rbrace~$[fm]&$\textrm{Im}\lbrace a_0\rbrace~$[fm]\\
      \hline\hline
      EFT$_\slashed{\pi}^\textrm{NLO}$&$8.48~$(input)&$7.5(6)$&$-2.6(\textrm{?})$\\
      \hline
      AV18&$7.57$&$7.81$&$-4.96$\\
      AV18+UIX&$8.43$&$7.62$&$-4.07$\\
      AV18+IL2&$8.48$&$7.63$&$-4.28$\\
      \hline
      \multirow{2}{*}{exp}&\multirow{2}{*}{$8.48$}&$7.456(20)^\textrm{\cite{exp-a0-huffman}}$&\\
      &&$7.370(58)^\textrm{\cite{exp-a0-zimmer}}$&\\
\hline
    \end{tabular}
\end{table}
The values for the recent IL2 3NI have been calculated for this work, employing methods
described in~\cite{mythesis,hmh-4he}.  The IL2 prediction for
$\textrm{Re}\lbrace a_0(^3\textrm{{\small He-n}})\rbrace$ is almost identical
to the UIX value and not plotted separately in fig.~\ref{fig.a0}. UIX and IL2
have parameters fitted to, amongst others, the triton and \mbox{4-Helium} binding
energies.  The observed deviation of the prediction of both models for
$a_0(^3\textrm{{\small He-n}})$ is therefore not a result of a deficiency of
the structure of the potential. The correlation between $B(t)$ and
$a_0(^3\textrm{{\small He-n}})$ supports the conjecture that this small deviation
can only be improved by the inclusion of higher order interactions.

Analogously to UIX, the IL2 model easily satisfies the criteria mentioned in sect.~\ref{ch.res.friar-line},
namely to reproduce low-energy NN observables and the triton binding energy
at least with a $10$\% accuracy, and therefore should with its prediction
for $a_0(^3\textrm{{\small He-n}})$ also lie within a $10$\% radius of the datum in fig.~\ref{fig.a0}.
This EFT prediction is confirmed here by explicit calculation.

Both experimental values for $a_0(^3\textrm{{\small He-n}})$ are included in the
predicted universality band, and therefore we cannot resolve the discrepancy
between the two measurements.
\subsection{Evidence for the absence of four-nucleon interactions at NLO}
We close with a comment on the implications of these results for the scaling
of four-nucleon interactions (4NI) in EFT$_\slashed{\pi}$ at LO and NLO.

Simplistic dimensional analysis suggests that a momentum-independent 4NI
enters at N$^4$LO, but the unusual renormalization of the 3NI may also promote
the 4NI to contribute at lower orders to ensure renormalizability.  By varying
their harsh cutoff between $800~$MeV and $2000~$MeV, Platter \textit{et al.}~\cite{platter-tjon} concluded
that a 4NI is not necessary at LO for cutoff-independence of the 4-Helium binding
energy.

In the present approach with a Gaussian regulator, the cutoff was varied for
the 4-Helium binding energy and for the 3-Helium-n scattering length from $400~$MeV to $700~$MeV.
This led to a correlation band near-identical with the ones mapped out when
using different NN potentials (in part also at different cutoffs) or different
3NIs. All results were converged  numerically. Moreover, we demonstrated that
the theoretical uncertainty for $B(\alpha)$ decreases from LO to NLO by a factor
consistent with the a-priori expansion parameter estimate $Q\approx\frac{1}{3}$. In both
cases, the physical datum lies well inside the NLO correlation band.

Let us assume now that a 4NI enters at NLO.  Our not accounting for it would
then have two effects: The results could be unstable against cutoff
variations, indicating that a 4NI is
necessary at NLO to renormalize EFT$_\slashed{\pi}$. We see no such effect, but are
aware that the cutoff might have to be varied beyond the window chosen in
this exploratory study. Secondly,  the theoretical accuracy of the results
would be reduced to LO, even if results independent of the cutoff can be
achieved, \textit{i.e.}, even if the cutoff can be removed to infinity without a
4NI. In other words, when one separately or combinedly varies the cutoff
or the 2-nucleon potential used or the 3NI strength, the different results should
spread in a corridor set by the size of LO corrections, \textit{i.e.}, $\lesssim 30$\%,
and not by the corridor of $\lesssim 10$\% expected in a NLO
calculation. We have demonstrated above for both observables that the residual
short-distance dependence of the observables is $\lesssim 10$\% of the central
value, taking as conservative estimate the combination of errors which occur
when varying EFT$_\slashed{\pi}$ at unphysically short distances: Different, phase
equivalent NN potentials; different 3NIs; different cutoffs. The correlation
bands and error estimates are thus on the quantitative level consistent with those
of a NLO calculation.

We also note in passing that  the physical datum lies well inside the NLO
correlation band for both four-nucleon observables. A residual 4NI not necessary for
renormalization but with unnaturally large coefficient can therefore be ruled out.

Our conclusion therefore is: There is strong evidence that 4NIs do not
enter at either LO or NLO in EFT$_\slashed{\pi}$. An indisputable criterion to assess
whether our interpretation is correct can be provided in a future study~\cite{kirscher2} of
the convergence pattern of the momentum-dependence of four-nucleon observables from zero
to the breakdown scale. In it, an observable
calculated up to order $Q^n$ must show a residual short-distance dependence on
compatible with $n$ powers of the typical low-energy momentum. This analysis
will not re-sum the effective range contributions into the 2-nucleon propagator,
but treat higher order effects in strict perturbation.
\section{Conclusions}\label{ch.concl}

The effective field theory formalism can be used to explain empirically found
correlations amongst few-nucleon observables like the Phillips- and Tjon line.
In these two cases, it relates the deviations from data to, first, an incomplete renormalization in the
3-nucleon sector, and second, to higher order interactions omitted in the course of
the EFT expansion. The theoretical uncertainty at every order of the calculation can be quantified
in this approach.
Numerical inaccuracies were demonstrated to be negligible in app.~\ref{ch.rrgm.numstab} and app.~\ref{ch.rrgm.numstab-2}.
Different results for different short-distance parameterizations therefore have
other origins.
The size of the expansion parameter $p_\text{typ}/\Lambda_b$
determines how fast an EFT expansion converges and decides its usefulness for
the calculation of an observable in a given system.
In heavier systems like \mbox{4-Helium}, the expansion parameter can \textit{a priori} be as large as $1$,
so they are border line. However, we find that next-to-leading order corrections to leading-order
results are still parametrically small in the four-nucleon system. This confirms a pattern
already seen in 2- and 3-nucleon systems at momenta which approach the \textit{a priori}
breakdown scale but where convergence is still found, see \textit{e.g.}~\cite{christl-ddis,bedaque-hgrie-tni,hgrie-zpara,rupak-npdgamma}.

The computational challenges of calculating few-nucleon observables are met in this work
by the Refined Resonating Group Method. It provides a versatile method
for bound- and scattering properties. Here, it was found highly economical with respect to
computer time when combined with the EFT$_\slashed{\pi}$ NN potentials derived in sect.~\ref{ch.potfit} for a
range of momentum cutoff values and with a full treatment of the Coulomb interaction.

For two correlations for which leading-order calculations exist, namely between the binding energy and charge radius
of the triton in sect.~\ref{ch.res.friar-line} and the binding energies of \mbox{4-Helium} and the triton in sect.~\ref{ch.res.tjon},
our coordinate space EFT$_\slashed{\pi}$ calculations at NLO report the expected improvement from LO to NLO
consistent with an expansion parameter $p_\text{typ}/\Lambda_b\approx\frac{1}{3}$. By that we demonstrated that
a consistent description of the $\alpha$-particle is possible at NLO in EFT$_\slashed{\pi}$.
In sect.~\ref{ch.res.t-3he}, we also report a correlation between the triton
binding energy and its difference to the $^3$He binding energy. As the
EFT$_\slashed{\pi}$ potential is at NLO iso-spin symmetric, this
model independent difference is attributed to Coulomb interactions only, which
are included in the RRGM. At the physical triton binding energy, this value
agrees well both in magnitude and uncertainty with estimates of
charge-symmetry breaking and Coulomb contributions to $^3$He binding.
In sect.~\ref{ch.res.a0}, a new correlation between the triton binding energy and the real part of the singlet S-wave scattering length
of \mbox{3-Helium-neutron} scattering similar to the Tjon line is also found. This, and the three aforementioned correlation bands, let us also conclude that no
four-body contact interaction is required to renormalize the system at next-to-leading order.
The position of all four bands, which represent universal properties of the 2-nucleon system,
was determined by fitting nine NN potentials differing at short-distances but with identical
long-distance behavior, by variations of the 3NI strengths, and by changing the cutoff.

Consistent with a basic tenet of EFT, namely model independence,
we also showed that the results of the phenomenological models AV18(+UIX/IL2), which share the
input of our EFT$_\slashed{\pi}$ NLO potentials, agree with their results within NLO accuracy.

Future work will utilize the relatively fast computations of four-nucleon observables, which result from the harmonic interplay
of the RRGM with EFT$_\slashed{\pi}^\textrm{NLO}$ potentials, in analyses of universal
properties of $A>4$ systems.
The Borromean halo nucleus, 6-Helium, is of special interest because EFT$_\slashed{\pi}$ calculations
in this system can also provide input for $\alpha N$ effective field theories~\cite{bertulani-halo}.
Furthermore, we are now equipped for a study of electro-weak interactions with heavier nuclei like
$^4$He($\gamma$,p) and $^4$He($\gamma$,n) using EFT$_\slashed{\pi}$ and the RRGM, to address both conflicting measurements
as well as theoretical calculations~\cite{shima,quagl,sand}. Valuable input for astrophysical calculations for the prediction
of light element abundances can also be provided by a calculation of reaction cross-sections for \textit{e.g.}
d(d,n)$^3$He and d(d,p)t at very low energies.

\begin{acknowledgments}
We thank M.C.~Birse, G.~Hager, H.W.~Hammer, C.~Pelissier, D.R.~Phillips, and G.~Wellein for helpful
discussions.
JK acknowledges the kind hospitality of the Dipartimento di Fisica of the
Universit\`a degli Studi di Trento and is in great debt to G.~Orlandini,
W.~Leidemann, and N.~Barnea for the many instructive conversations and
critical comments.
The computational resources for this work were in part provided by
the RRZE of Universit\"at Erlangen-N\"urnberg.
HWG is also grateful for the hospitality of the Institut f\"ur Theoretische
Physik III at Universit\"at Erlangen-N\"urnberg, of the Institut f\"ur
Theoretische Physik (T39) at TU M\"unchen, of the Nuclear Experiment group
of the Institut Laue-Langevin (Grenoble, France), and to the organizers and
participants of the workshop ``Bound States and Resonances in EFTs'' at the
ECT* (Trento, Italy) for stimulating discussions.
Finally, we are indebted to the referees for encouraging major improvements and
clarifications.
This work was supported in part by the CAREER-grant PHY-0645498 of the US
National Science Foundation and by the US Department of Energy grants
DE-FG02-95ER-40907 and DE-FG02-97ER-41019.
\end{acknowledgments}
\appendix
\section{Numerical stability: two-nucleon sector}\label{ch.rrgm.numstab}
In the analysis of sect.~\ref{ch.res}, various methods are used to ascertain higher order effects and
provide reliable theoretical uncertainties of our NLO calculations. It is
therefore imperative to ensure that purely numerical inaccuracies of the variational method are not a
significant source of error.

When increasing the model space, two obstacles have to be considered: first,
numerical linear dependences amongst the basis vectors; and second, too broad
distortion channels.  Both issues lead to unstable results and occur
especially when the dimension of the model space becomes large.  The bulk
behavior of the phase shifts in the 4-Helium system
was stable in that respect, but the prediction for the
$^3$He-n spin-zero S-wave scattering length was much more sensitive to the
addition of basis vectors.  This is a consequence of too broad distortion
channels, which is associated with the
expansion of the Coulomb functions.  We illustrate this point here as a purely
numerical issue, anticipating results from sect.~\ref{ch.res}. All integrals
needed to calculate $a_{\lambda j}$ and $b_{\lambda jm}$ have a support only
for $R_j<R_{\textrm{max}}$, where $R_{\textrm{max}}$ is set by the
size of the largest fragment.  When the Coulomb functions
are expanded in Gaussians, this inner region is weighted more at the expense
of a less accurate fit for larger separations.  This poses no problem if broad
Gaussians are used for the distortion channels, \textit{i.e.},
$\omega_{jm}<\omega_{\textrm{min}}(R_{\textrm{max}})$, because the results are
only affected by the values of the Coulomb function in the outer region,
$R_j>R_{\textrm{max}}$. In sect.~\ref{ch.res}, potentials are presented which
yield relatively weakly bound $^3$H nuclei with more extended wave functions
(see fig.~\ref{fig.rch}). For those interactions, distortion channels with
broader width parameters had to be included in order to form the triton
cluster inside of the \mbox{4-Helium} nucleus and
reach convergence for the \mbox{4-Helium} binding energy. If, on the other hand, too
many of those broad vectors are included, the phase shifts and especially the
scattering length $a_0(^3\textrm{{\small He-n}})$ become unstable.

The NN model space used to fit the LECs (see sect.~\ref{ch.potfit}) is a key
factor. If the scattering and bound states of a given set of parameters cannot
be expanded accurately, the resultant potential will carry a non-negligible
model space dependence.  To minimize this dependence, a large two-body model
space was chosen with a set of width parameters which cover an interval
corresponding to the expected extention of the 2-nucleon wave
function. Twenty width parameters in the range
$\omega_{jm}\in[0.001\,\textrm{fm}^{-2};130\,\textrm{fm}^{-2}]$ (called set
$w_{120}$~\cite{reiss-ay-widths}) were used for the NN scattering states, and
the same set for the widths $\gamma_{dk}$ in eq.~(\ref{eq.bsbv}) for the S-
and D-wave components of the deuteron bound state. Completeness of this set
was defined to be sufficient by comparing the results (see
table~\ref{tab.pot-nn-obs}) to those of another set of 20 parameters, where each
inverse width of $w_{120}$ is divided by ten, resulting in broader
widths. Hence, all widths of this set $w_{12}$ lie in the interval
$[0.0001\,\textrm{fm}^{-2};13\,\textrm{fm}^{-2}]$.  For the triton and
\mbox{3-Helium} wave functions, a previously mentioned set of 224 vectors taken
from~\cite{hmh-4he} spanned the model space. It
contains all possible couplings with total angular momentum $L\leq 2$
and 142 different width parameters.  To model the deuteron
fragment in the \mbox{4-Helium} calculation, its model space was reduced to 6 S-wave
and 3 D-wave widths for each potential separately, employing a genetic
algorithm which optimizes the width parameters of the basis
vectors~\cite{genalg}. These sets are named $w_{63}$ in what follows.
These optimized deuteron width sets for narrower
potentials with larger cutoff values also contained narrower widths, in line
with the need to model wave functions with a more complicated short-distance
structure.

The expansion of the wave function in terms of Gaussians had significant
impact on the choice of the regulator. First, a cutoff momentum space regulator
$f(\vec{q})=\text{exp}\left(-\vec{q}^2/\Lambda^2\right)$, leads to a Gaussian radial
dependence as shown below in eq.~(\ref{eq.pot-coord}). Hence, all radial
dependences of the potential are Gaussians and can directly be implemented
into the RRGM. Second, with its functional form fixed, the magnitude of $\Lambda$
fully specifies the regulator and is also influenced by the RRGM. A lower bound is set by the breakdown
scale $\Lambda_b\approx m_\pi$ of EFT$_\slashed{\pi}$, since modes which lie
in the range of applicability of EFT$_\slashed{\pi}$ should not be
suppressed. On the other hand, an
upper bound $\Lambda_t$ for the cutoff is set by the numerics of the RRGM. Large cutoff
values correspond to narrower Gaussian potentials in coordinate space, and
hence the RRGM model space must include narrower Gaussians as well to expand
the corresponding, more localized NN bound state accurately, see below. Clearly,
$\Lambda$ cannot be chosen considerably larger than the inverse of the
narrowest width in which one expands. A LO calculation in the deuteron
channel, \textit{i.e.} $V(\vec{r})=C_t\text{exp}\left(-\Lambda^2\vec{r}^2/4\right)$, was carried out to
illustrate this point.
\begin{figure}
  \includegraphics[width=\columnwidth]{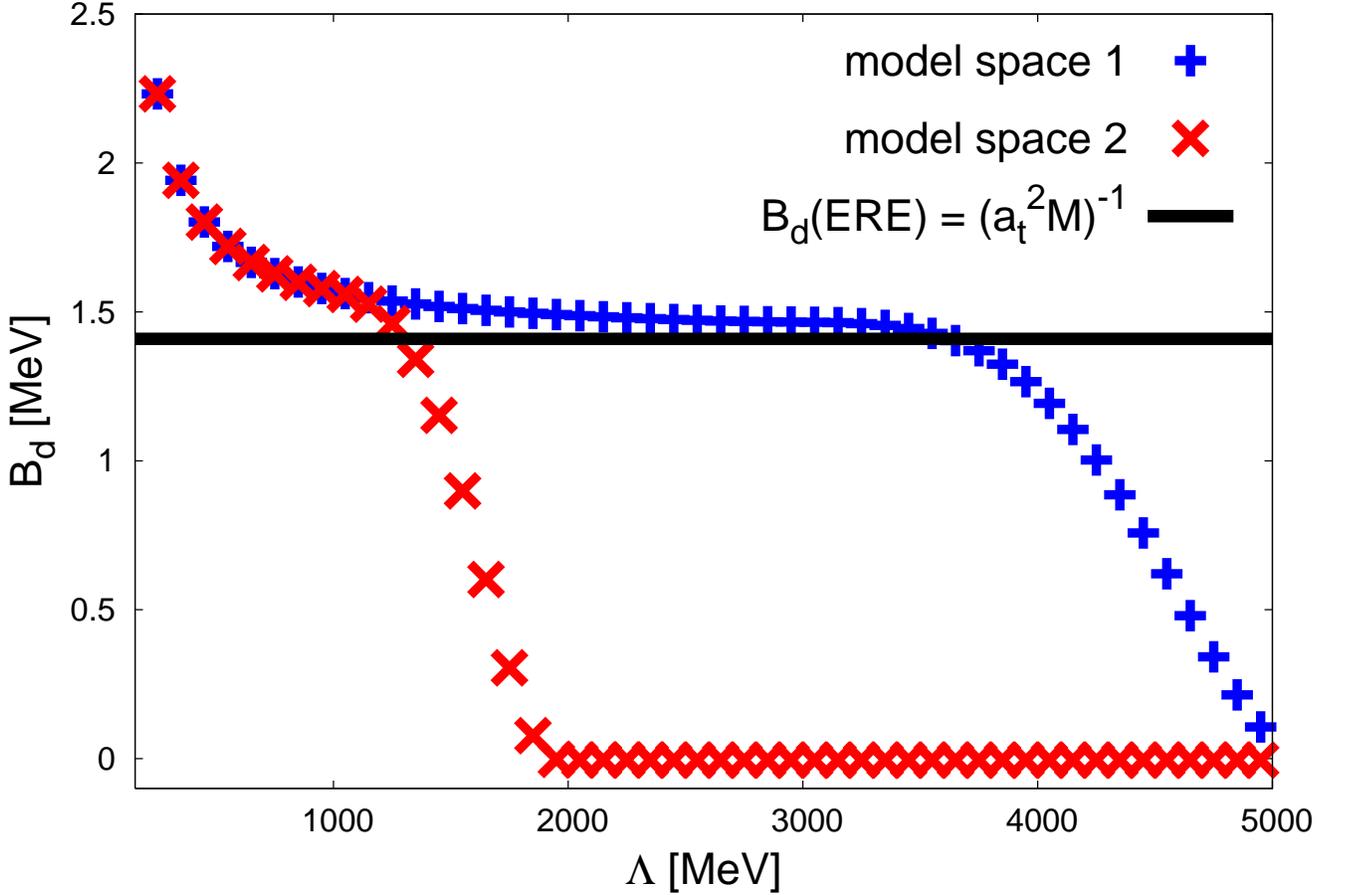}
  \caption{\label{fig.bd-cutoff} Binding energy $B_d$ of the deuteron in
    leading-order EFT$_\slashed{\pi}$ as a function of the Gaussian momentum
    cutoff $\Lambda$ for two different RRGM model spaces. The space $w_{120}$
    contains narrower width parameters, and hence can be used for the
    potentials corresponding to larger values of $\Lambda$. Using the triplet
    neutron-proton scattering length as experimental input, the effective
    range formula predicts $B_d=1.41~$MeV.}
\end{figure}
The low-energy constant $C_t$ was fitted to reproduce the experimental neutron-proton triplet scattering
length $a_t^\textrm{exp}$ by numerically solving
the two-body Schr\"odinger equation with MATHEMATICA. The scattering
length was calculated from the phase shift at ${E_\textrm{cm}=10^{-4}~}$MeV, and the resulting
coefficient $C_t$ was then fed into the RRGM code.
The RRGM result for the deuteron binding
energy is plotted against $\Lambda$ in fig.~\ref{fig.bd-cutoff}. The LO
prediction of the effective range expansion, $B(d)=1.4096~$MeV from
$a_t^\textrm{exp}=0.02748~\textrm{MeV}^{-1}$~\cite{exp-anp}, is reproduced
in both considered model spaces consistently in the RRGM calculation up to the
cutoff $\Lambda_t$.  Above this threshold, $\Lambda_t^{w12}\approx 1.4~$GeV and
$\Lambda_t^{w120}\approx 3.6~$GeV, the RRGM solution becomes
strongly cutoff-dependent because the respective model space is
insufficient to expand a bound state corresponding to such narrow potentials.
We explain this as follows.
The radial ground state wave function reaches a maximum around $r_p\approx \Lambda^{-1}$, before decaying exponentially
outside the interaction region. Most economically, this peak can be approximated
by a sum of two Gaussians with widths comparable to $r_p$.
For the expansion of the exponential decay, broader Gaussians suffice.
If the model space does not contain Gaussians of width $r_p$, an approximation of the increasingly steep
rise of the wave function at zero accompanying an increasing $\Lambda$ will eventually fail.
Hence, bound states can no longer be expanded in the model space for
potentials of shorter range than the narrowest Gaussian basis state.
In the case shown in fig.~\ref{fig.bd-cutoff}, relating the narrowest Gaussian widths $\omega_{\textrm{max}}$
of the two model spaces, $w_{12}$ and $w_{120}$, via the regulator $\text{exp}\left(-\Lambda^2\vec{r}^2/4\right)$
to a cutoff $\Lambda_t=2\sqrt{\omega_{\textrm{max}}}$,
one expects the ground state in the respective model space to become unbound for cutoffs larger than
$\Lambda^{w12}_t\approx 1.4~$GeV and $\Lambda^{w120}_t\approx 4.5~$GeV. The thresholds suggested
in fig.~\ref{fig.bd-cutoff} by a drop of the deuteron binding energy are of this expected magnitude.
This lead to the conjecture that the $w_{120}$ RRGM model space is appropriate
for calculations with cutoff values in a range $150~\text{MeV}\lesssim\Lambda\lesssim 3~\text{GeV}$.

\begin{figure}
  \includegraphics[width=\columnwidth]{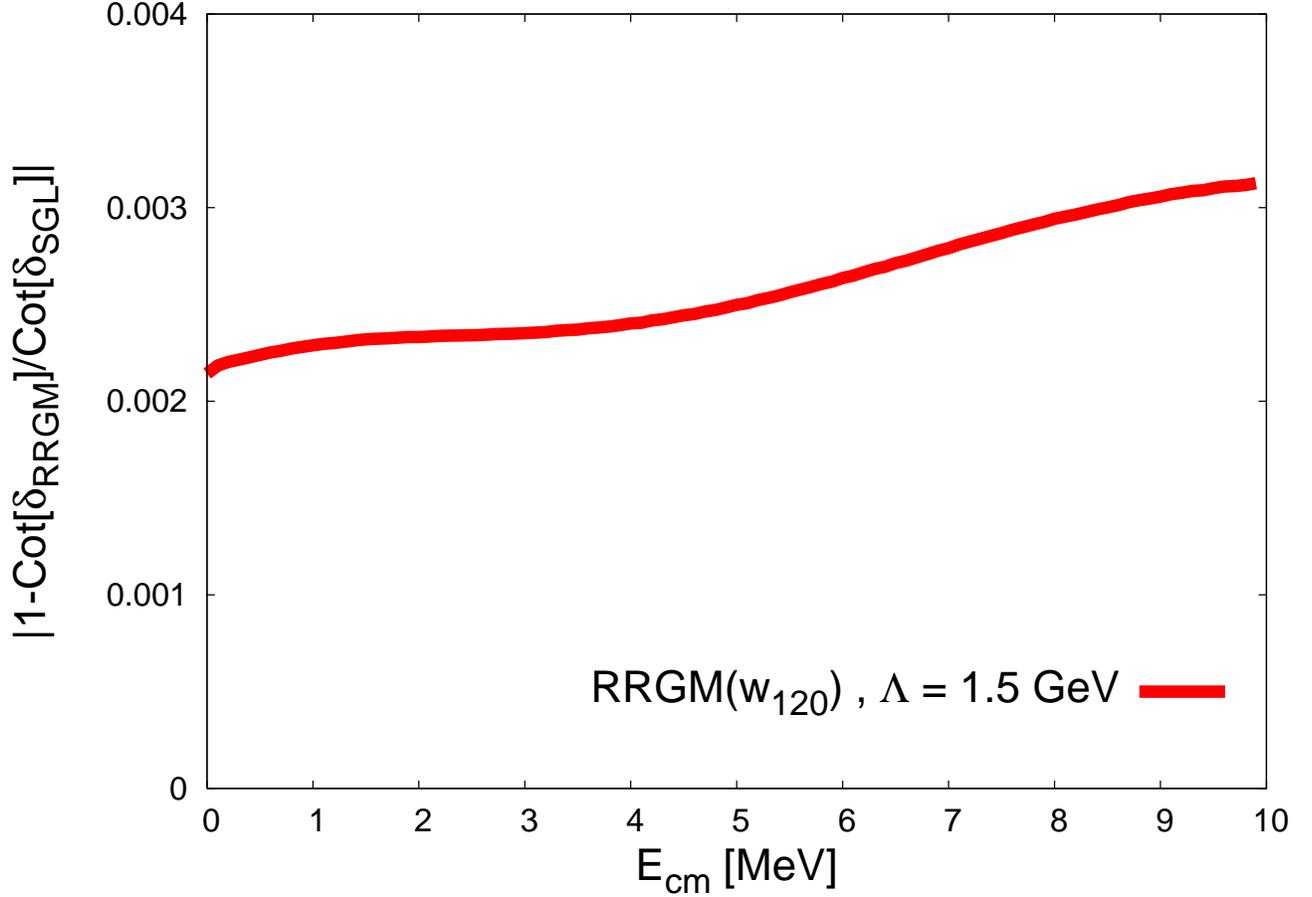}
  \caption{\label{fig.sgl-kett} Relative deviation between the $^3S_1$ NN phase shift results obtained
with the RRGM and a numerical integration of the Schr\"odinger equation. A leading-order potential with one parameter fit to $a_t$ and a
cutoff $\Lambda=1.5~$GeV was used. Notice the scale on the ordinate.}
\end{figure}

A comparison of the phase shifts $\delta(^3S_1)$ below $E_\textrm{cm}=10~$MeV
resulting from an RRGM calculation on the one side, and of a numerical
integration of the Schr\"odinger equation on the other strongly supports this.
The results shown in fig.~\ref{fig.sgl-kett} of a potential
with a cutoff of $1.5~$GeV differ by less than $0.3$\%.
As long as the cutoff was kept below the threshold $\Lambda_t$, the relative difference was always found to be of that order.
For cutoff values approaching or surpassing $\Lambda_t$, the discrepancy increased considerably indicating the
predicted failure of the specific RRGM model space.

For the NLO potentials, $\Lambda$ was taken from the interval between
$400~\textrm{MeV}$ and $1~\textrm{GeV}$, in which $w_{12}$ and
$w_{120}$ yielded the same deuteron binding energy.
Therefore, it is reasonable to assume that both model spaces are large
enough to expand the scattering and bound states of the NLO potentials,
so that the model space dependence of the potentials should be minimal.
Differences between results using different potentials and short-distance
physics do therefore not stem from numerical inaccuracies.

\section{Numerical stability: four-nucleon scattering}\label{ch.rrgm.numstab-2}
\begin{figure}
  \includegraphics[width=\columnwidth]{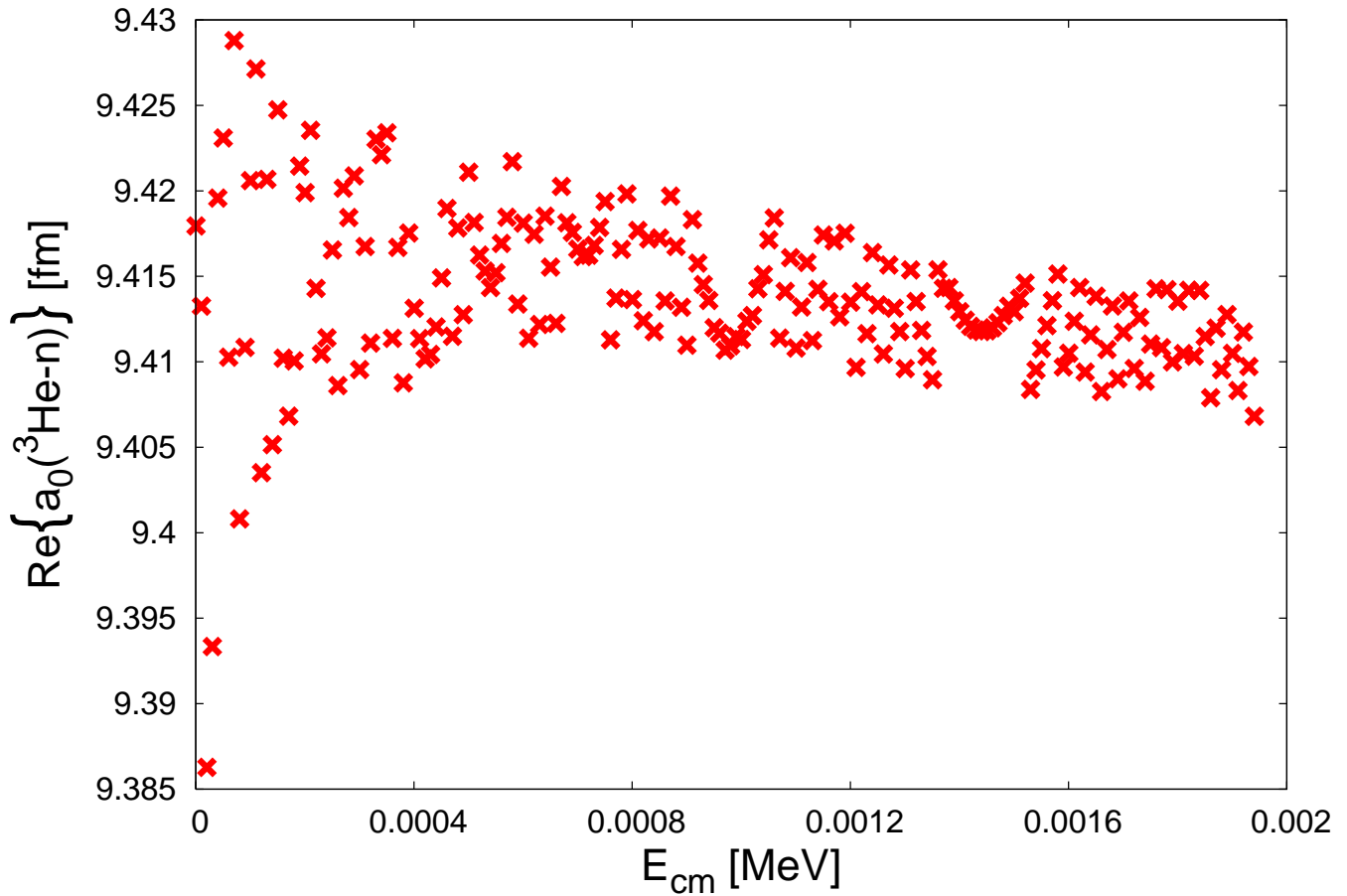}
  \caption{\label{fig.a0ofe} The real part of the spin singlet, S-wave
    scattering length for elastic $^3$He-n scattering as a function of the
    center of mass energy above the $^3$He-n threshold at which
    $a_0(^3\textrm{{\small He-n}})$ is calculated from
    eq.~(\ref{eq.scatt-length}). Notice the scale on the vertical axis. Taking
    the central value of the band as the actual value, its decrease from
    200~eV to 2~keV is due to effective range corrections, while the band is a
    result of numerical noise. $^8V_\slashed{\pi}$ was used for this plot,
    with the other $V_\slashed{\pi}$ potentials showing similar behavior
    with respect to the size of the higher order corrections and numerical
    fluctuations.}
\end{figure}
To gauge the size of the error introduced by numerical uncertainties
from diagonalizing the Hamiltonian for the calculations with zero 3NI, $a_0(^3\textrm{{\small He-n}})$ was
calculated over a wider range of energies, $E_\textrm{cm}\leq 2~$keV.
Figure~\ref{fig.a0ofe} shows $\textrm{Re}\lbrace a_0(^3\textrm{{\small He-n}})\rbrace$
as a function of the matching energy at which it is
calculated for the potential $^8V_\slashed{\pi}$.  The fluctuations resulting
in the band-like shape with a width of about $0.015$~fm out of
  $a_0(^3\textrm{{\small He-n}})\approx9.415$~fm are due to uncertainties
associated with the diagonalization of numerically singular matrices.  The
error due to effective range corrections can be estimated by observing the
difference,
\begin{equation}\label{eq.a0corr}
  |\textrm{Re}\lbrace a_0(200~\textrm{eV})\rbrace-\textrm{Re}\lbrace a_0(2~\textrm{keV})\rbrace|\approx 0.01~\textrm{fm}\;\;\;.
\end{equation}
We therefore conclude that the numerical precision of our result is about
$1\%$ and is considerably smaller than the theoretical accuracy of $\sim10\%$
attributed to a NLO calculation. For AV18, the same analysis produced stable
results with respect to numerical fluctuations, \textit{i.e.} no visible band
as in fig.~\ref{fig.a0ofe} is found. This is attributed to the model
space used for AV18, which was optimized for this potential and purged of
states with large but mutually canceling overlap with the ground state.
As mentioned at the beginning of this paragraph, this analysis
applies to values obtained in an optimized model space only.

\section{Computing time requirements}\label{ch.res.comp}
\begin{table}
\caption{\label{tab.cputime}Runtime comparison between RRGM calculations with EFT and phenomenological potentials. dim(rad-dep) is
the number of Gaussians needed to expand the radial dependences, and dim(ms) is proportional to the dimension of the model space.
The parallel computations were performed on IA32 Xeon $2.66~$GHz nodes of the
RRZE in Erlangen, while for serial code, an E6750 architecture at $2.66~$GHz was used.}
    \begin{tabular}{clcc|r}
Observable & Potential & dim(rad-dep) & dim(ms) & Runtime\\
&&&& \small{[CPUh]}\\
\hline\hline
$B(t)$ &AV18&$44$&\multirow{2}{*}{67}&$0.04$\\
      &UIX(parallel)&$204$&&$0.17$\\
      &EFT $^1V_\slashed{\pi}$&$9$&67&$<0.01$\\
      &$^1V_\slashed{\pi}$+3NI&$1$&&$<0.01$\\
\hline
$B(\alpha)$ &AV18(parallel)&$44$&\multirow{2}{*}{324}&$2.6$\\
      &UIX(parallel)&$204$&&$6.8$\\
      &EFT $^1V_\slashed{\pi}$&$9$&550&$0.45$\\
      &$^1V_\slashed{\pi}$+3NI&$1$&&$0.03$\\
\hline
$a_0(^3\textrm{{\small He-n}})$ &AV18(parallel)&$44$&\multirow{2}{*}{324}&$10.4$\\
      &UIX(parallel)&$204$&&$27.1$\\
      &EFT $^1V_\slashed{\pi}$&$9$&550&$1.8$\\
\hline
    \end{tabular}
\end{table}
In table~\ref{tab.cputime}, the computer time, normalized to one CPU, of the EFT$_\slashed{\pi}$ and the AV18(+UIX)
calculations is compared. We display the time used for the build-up of the Hamilton matrix, which surpasses
by at least an order of magnitude the
time needed for the other stages, namely the calculation of the spin- and coordinate space matrix elements and the
diagonalization of the Hamilton matrix. The EFT potentials allow for faster calculations
because we choose the regulator of the EFT potentials such that no expansion of the radial dependences is necessary. Hence, only one matrix element
is evaluated for a specific operator. For the same operator which has a different radial dependence in AV18, one has to
calculate as many matrix elements as Gaussian basis functions are necessary to approximate this dependence accurately.
The calculational time per term in the Gaussian expansion is dominated by the operator structure only,
and is found to be comparable for AV18 and the EFT potentials.
This resembles the already stressed occurrence of all AV18 operators in the EFT potentials considered here.
The central 3NI is equally inexpensive with respect to computation time for systems with $A\leq 4$ as the NN potentials.
We plan to invest the computer time saved by the convenient choice of the regulator of the EFT$_\slashed{\pi}$ interaction
in calculating heavier nuclear systems~\cite{kirscher2}. For those systems, the model space and the number of
possible spin and coordinate spaces coupling schemas increases the number of matrix elements to be summed up
considerably. Having reduced the number of matrix elements from a Gaussian expansion of the potential, therefore,
allows for more refined and larger model spaces.
\bibliographystyle{unsrt}
\bibliography{nlo-piless-rrgm-feas-after-ref}

\begin{thebibliography}{10}

\bibitem{eft-rev-vK}
U.~van Kolck.
\newblock {\em Prog. Part. Nucl. Phys.}, 43:337--418, 1999.

\bibitem{eft-rev-bb}
S.R. Beane, P.F. Bedaque, W.C. Haxton, D.R. Phillips, and M.J. Savage.
\newblock {From hadrons to nuclei: Crossing the border}.
\newblock In M.~Shifman, editor, {\em At the frontier of particle physics},
  volume~1, pages 133--269, 2000.
\newblock arXiv:nucl-th/0008064.

\bibitem{eft-rev}
P.F. Bedaque and U.~van Kolck.
\newblock {\em Ann. Rev. Nucl. Part. Sci.}, 52:339--396, 2002.

\bibitem{eft-rev-platter}
L.~Platter.
\newblock {\em Few Body Syst.}, 46:139--171, 2009.

\bibitem{brat-hammer}
H.W. Hammer and E.~Braaten.
\newblock {\em Phys. Rept.}, 428:259--390, 2006.

\bibitem{mod-rev-nucl-epel}
E.~Epelbaum, H.W. Hammer, and U.G. Mei{\ss}ner.
\newblock {\em Rev. Mod. Phys.}, 81:1773--1825, 2009.

\bibitem{xpt-epelbaum}
E.~Epelbaum.
\newblock {\em Nucl. Phys. A}, 737:43--51, 2004.

\bibitem{epel-chipt-rev}
E.~Epelbaum.
\newblock 2010.
\newblock arXiv:nucl-th/1001.3229.

\bibitem{rupak-npdgamma}
G.~Rupak.
\newblock {\em Nucl. Phys. A}, 678:405, 2000.

\bibitem{platter-tjon}
L.~Platter, H.W. Hammer, and U.G. Mei{\ss}ner.
\newblock {\em Phys. Lett. B}, 607:254--258, 2005.

\bibitem{vkol-ncsm-li}
I.~Stetcu, B.R. Barrett, and U.~van Kolck.
\newblock {\em Phys. Lett. B}, 653:358--362, 2007.

\bibitem{tjon}
J.A. Tjon.
\newblock {\em Phys. Lett. B}, 56:217, 1975.

\bibitem{phillips}
A.C. Phillips.
\newblock {\em Nucl. Phys. A}, 107:209, 1968.

\bibitem{efi-plat-hamm}
H.W. Hammer and L.~Platter.
\newblock 2010.
\newblock arXiv:nucl-th/1001.1981.

\bibitem{efimov-spect}
V.~Efimov.
\newblock {\em Sov. J. Nucl. Phys.}, 12:589, 1971.

\bibitem{bedaque-tni}
P.F. Bedaque, H.W. Hammer, and U.~van Kolck.
\newblock {\em Phys. Rev. Lett.}, 82:463--467, 1999.

\bibitem{kharchenko}
V.F. Kharchenko.
\newblock {\em Sov. J. Nucl. Phys.}, 16:173, 1973.

\bibitem{hmh-rrgm}
H.M. Hofmann.
\newblock In L.~S. Ferreira, A.~C. Fonseca, and L.~Streit, editors, {\em
  Proceedings of Models and Methods in Few-Body Physics, Lisboa, Portugal},
  page 243, 1986.

\bibitem{eft-rev-phill}
D.R. Phillips.
\newblock {\em Czech. J. Phys.}, 52:B49, 2002.

\bibitem{eft-ord-ray-vkol}
C.~Ordonez, L.~Ray, and U.~van Kolck.
\newblock {\em Phys. Rev. Lett.}, 72:1982, 1994.

\bibitem{bedaque-tni-boson}
P.F. Bedaque, H.W. Hammer, and U.~van Kolck.
\newblock {\em Nucl. Phys. A}, 646:444--466, 1999.

\bibitem{bedaque-hgrie-tni}
P.F. Bedaque, G.~Rupak, H.W. Grie{\ss}hammer, and H.W. Hammer.
\newblock {\em Nucl. Phys. A}, 714:589--610, 2003.

\bibitem{ha-meh-nd-scatt}
H.W. Hammer and T.~Mehen.
\newblock {\em Phys. Lett.}, B516:353--361, 2001.

\bibitem{platter-NNLO}
L.~Platter.
\newblock {\em Phys. Rev. C}, 74:037001, 2006.

\bibitem{platt-phill-NNLO}
L.~Platter and D.R. Phillips.
\newblock {\em FewBodySyst.}, 40:35--55, 2006.

\bibitem{lepage}
G.P. Lepage.
\newblock {How to renormalize the Schroedinger equation}.
\newblock 1997.
\newblock lectures given at 9th Jorge Andre Swieca Summer School: Particles and
  Fields, Sao Paulo, Brazil, 16-28 Feb 1997. arXiv:nucl-th/9706029.

\bibitem{christl-ddis}
S.~Christlmeier and H.W. Grie{\ss}hammer.
\newblock {\em Phys. Rev. C}, 77:064001, 2008.

\bibitem{hgrie-zpara}
H.W. Grie{\ss}hammer.
\newblock {\em Nucl. Phys. A}, 744:192--226, 2004.

\bibitem{wein-chi-nucl}
S.~Weinberg.
\newblock {\em Phys. Lett. B}, 251:2, 1990.

\bibitem{beane-savage-rearr}
S.R. Beane and M.J. Savage.
\newblock {\em Nucl. Phys. A}, 694:511--524, 2001.

\bibitem{wign-bound-phil}
D.R. Phillips and T.D. Cohen.
\newblock {\em Phys. Lett.}, B390:7--12, 1997.

\bibitem{edmonds}
A.R. Edmonds.
\newblock {\em Angular Momentum in Quantum Mechanics}.
\newblock Princton University Press, 1996.

\bibitem{newton}
R.G. Newton.
\newblock {\em Scattering Theory of Waves and Particles}.
\newblock Dover Publications, 2002.
\newblock ch. 14.6.

\bibitem{genalg}
C.~Winkler and H.M. Hofmann.
\newblock {\em Phys. Rev. C}, 55:684--687, 1997.

\bibitem{av18}
R.B. Wiringa, V.G.J. Stoks, and R.~Schiavilla.
\newblock {\em Phys. Rev. C}, 51:38--51, 1995.

\bibitem{bonn}
R.~Machleidt, K.~Holinde, and C.~Elster.
\newblock {\em Phys. Rept.}, 149:1--89, 1987.

\bibitem{exp-deut}
G.L. Greene, E.G. Kessler, R.D. Deslattes, and H.~B\"orner.
\newblock {\em Phys. Rev. Lett.}, 56(8):819--822, Feb 1986.

\bibitem{exp-anp}
O.~Dumbrajs, R.~Koch, H.~Pilkuhn, G.C. Oades, H.~Behrens, J.J. de~Swart, and
  P.~Kroll.
\newblock {\em Nucl. Phys. B}, 216(2):277 -- 335, 1983.

\bibitem{nijm}
V.G.J. Stoks, R.A.M. Klomp, M.C.M. Rentmeester, and J.J. de~Swart.
\newblock {\em Phys. Rev. C}, 48:792--815, 1993.

\bibitem{nnon}
Nijmegen PWA online database, accessed 11/2008, http://nn-online.org/.

\bibitem{platter-rch}
L.~Platter and H.W. Hammer.
\newblock {\em Nucl. Phys. A}, 766:132--141, 2006.

\bibitem{exp-t-rch}
D.R. Tilley, H.R. Weller, and H.H. Hasan.
\newblock {\em Nucl. Phys. A}, 474(1):1 -- 60, 1987.

\bibitem{exp-bt}
A.H. Wapstra and G.~Audi.
\newblock {\em Nucl. Phys. A}, 432(1):1 -- 54, 1985.

\bibitem{rch-pots-1}
G.L. Payne, J.L. Friar, B.F. Gibson, and I.R. Afnan.
\newblock {\em Phys. Rev. C}, 22(2):823--831, 1980.

\bibitem{rch-pots-2}
C.R. Chen, G.L. Payne, J.L. Friar, and B.F. Gibson.
\newblock {\em Phys. Rev. C}, 31(6):2266--2273, 1985.

\bibitem{friar-rch}
J.L. Friar, B.F. Gibson, C.R. Chen, and G.L. Payne.
\newblock {\em Phys. Lett. B}, 161:241, 1985.

\bibitem{uix}
B.S. Pudliner, V.R. Pandharipande, J.~Carlson, S.C. Pieper, and R.B. Wiringa.
\newblock {\em Phys. Rev. C}, 56:1720--1750, 1997.

\bibitem{pest-pot}
G.H. Berthold, A.~Stadler, and H.~Zankel.
\newblock {\em Phys. Rev. C}, 38(1):444--448, 1988.

\bibitem{csb-tni}
J.L. Friar, G.L. Payne, and U.~van Kolck.
\newblock {\em Phys. Rev. C}, 71:024003, 2005.

\bibitem{opper-csb}
G.A. Miller, A.K. Opper, and E.J. Stephenson.
\newblock {\em Ann. Rev. Nucl. Part. Sci.}, 56:253--292, 2006.

\bibitem{exp-ba}
S.~Fiarman and W.E. Meyerhof.
\newblock {\em Nucl. Phys. A}, 206(1):1 -- 64, 1973.

\bibitem{mythesis}
J.~Kirscher.
\newblock Diploma thesis {FAU Erlangen}, 2006.
\newblock {http://theorie3.physik.uni-erlangen.de/theses/data/Dip-2006-01.pdf}.

\bibitem{delt-port-n3h}
A.~Deltuva and A.C. Fonseca.
\newblock {\em Phys. Rev. C}, 75:014005, 2007.

\bibitem{hmh-4he}
H.M. Hofmann and G.M. Hale.
\newblock {\em Phys. Rev. C}, 77:044002, 2008.

\bibitem{ill}
S.C. Pieper, V.R. Pandharipande, R.B. Wiringa, and J.~Carlson.
\newblock {\em Phys. Rev. C}, 64:014001, 2001.

\bibitem{exp-a0-huffman}
P.R. Huffman et~al.
\newblock {\em Phys. Rev. C}, 70:014004, 2004.

\bibitem{exp-a0-zimmer}
O.~Zimmer, G.~Ehlers, B.~Farago, H.~Humblot, W.~Ketter, and R.~Scherm.
\newblock {\em EPJdirect}, A1:1--28, 2002.

\bibitem{kirscher2}
J.~Kirscher, H.W. Grie{\ss}hammer, and H.M. Hofmann.
\newblock forthcoming.

\bibitem{bertulani-halo}
C.A. Bertulani, H.W. Hammer, and U.~Van Kolck.
\newblock {\em Nucl. Phys. A}, 712:37--58, 2002.

\bibitem{shima}
T.~Shima et~al.
\newblock {\em Phys. Rev. C}, 72:044004, 2005.

\bibitem{quagl}
S.~Quaglioni, W.~Leidemann, G.~Orlandini, N.~Barnea, and V.D. Efros.
\newblock {\em Phys. Rev. C}, 69:044002, 2004.

\bibitem{sand}
W.~Sandhas, W.~Schadow, G.~Ellerkmann, L.L. Howell, and S.A. Sofianos.
\newblock {\em Nucl. Phys. A}, 631:210c--229c, 1998.

\bibitem{reiss-ay-widths}
C.~Reiss and H.M. Hofmann.
\newblock {\em Nucl. Phys. A}, 716:107--119, 2003.

\end{thebibliography}
\end{document}